\documentclass[aps,prb,onecolumn,nofootinbib,citeautoscript,10pt]{revtex4-2}  
\usepackage{amsmath,amssymb,bm} 
\usepackage{graphicx,comment}

\usepackage[tight]{subfigure} 

\usepackage{color} 
\usepackage[papersize={8.5in,11in}]{geometry}
\usepackage[colorlinks=true]{hyperref}
\hypersetup{
bookmarks=true,         
unicode=false,          
pdftoolbar=true,        
pdfmenubar=true,        
pdffitwindow=false,     
pdfstartview={FitH},    
pdfkeywords={keyword1} {key2} {key3}, 
pdfnewwindow=true,      
colorlinks=true,       
linkcolor=magenta, 
citecolor=blue,        
filecolor=magenta,      
urlcolor=blue           
} 

\usepackage{graphicx}
\usepackage{dcolumn}
\usepackage{color}
\usepackage{amssymb,amsmath}
\usepackage{tabularx,graphicx}
\usepackage{epstopdf}
\usepackage{latexsym}
\usepackage{colortbl}
\usepackage{psfrag}
\usepackage{bbm,bm,array,physics}
\usepackage{dsfont}

\def \nn{\nonumber \\}

\begin{document}
\title{Electric and thermoelectric response for Weyl and multi-Weyl semimetals in planar Hall configurations including the effects of strain}

\author{Rahul Ghosh$^{1,2}$}
\author{Ipsita Mandal$^{1}$}

\affiliation{$^1$Department of Physics, Shiv Nadar Institution of Eminence (SNIoE), Gautam Buddha Nagar, Uttar Pradesh 201314, India\\
$^2$Dynamics Lab, Department of Chemistry, Indian Institute of Technology Delhi, New Delhi 110016, India}

\begin{abstract} 
We investigate the response tensors in planar Hall (or planar thermal Hall) configurations such that a three-dimensional Weyl or multi-Weyl semimetal is subjected to the influence of an electric field $\mathbf E $ (or temperature gradient $\nabla_{\mathbf r } T$) and an effective magnetic field $\mathbf B^{\rm tot}$, which are oriented at a generic angle with respect to each other. The effective magnetic field consists of two parts --- (a) an actual/physical magnetic field $\mathbf B $, and (b) an emergent magnetic field $\mathbf B_5 $ which quantifies the elastic deformations of the sample. $\mathbf B_5 $ is an axial pseudomagnetic field because it couples to conjugate nodal points with opposite chiralities with opposite signs. We study the interplay of the orientations of these two components of $\mathbf B^{\rm tot}$ with respect to the direction of the electric field (or temperature gradient) and elucidate how it affects the characteristics involving the chirality of the node. Additionally, we show that the magnitude and sharpness of the conductivity tensor profiles strongly depend on the value of the topological charge at the node in question.
\end{abstract}

\maketitle

\tableofcontents

\section{Introduction}
\label{sec:intro}

There has been a surge of investigations of the transport properties of semimetallic systems which harbour two or more band-crossing points in the Brillouin zone (BZ), where the density of states goes to zero. Among the well-known three-dimensional (3d) semimetals are the Weyl semimetals (WSMs) \cite{burkov11_weyl,yan17_topological} and the multi-Weyl semimetals (mWSMs) \cite{bernevig,bernevig2,dantas18_magnetotransport}, whose bandstructures exhibit nontrivial topological features, as each nodal point can be considered as a source or sink of the Berry flux. In other words, a nodal point can be thought of as an analogue of a magnetic monopole in the momentum space, with the monopole charge giving rise to a nonzero Chern number (arising from the Berry connection). The Nielsen-Ninomiya theorem \cite{nielsen} imposes the condition that the nodes come in pairs such that the two nodes carry Chern numbers $\pm J$, which are equal in magnitude but opposite in signs. The sign of the monopole charge is often referred to as the chirality of the corresponding node. The Chern numbers of Weyl, double-Weyl (e.g., $\mathrm{HgCr_2Se_4}$~\cite{Gang2011} and $\mathrm{SrSi_2}$~\cite{hasan_mweyl16}), and triple-Weyl nodes (e.g., transition-metal monochalcogenides~\cite{liu2017predicted}) are 
$\pm 1$, $\pm 2$, and $\pm 3$, respectively. 

\begin{figure*}[t]
\includegraphics[width=0.5 \linewidth]{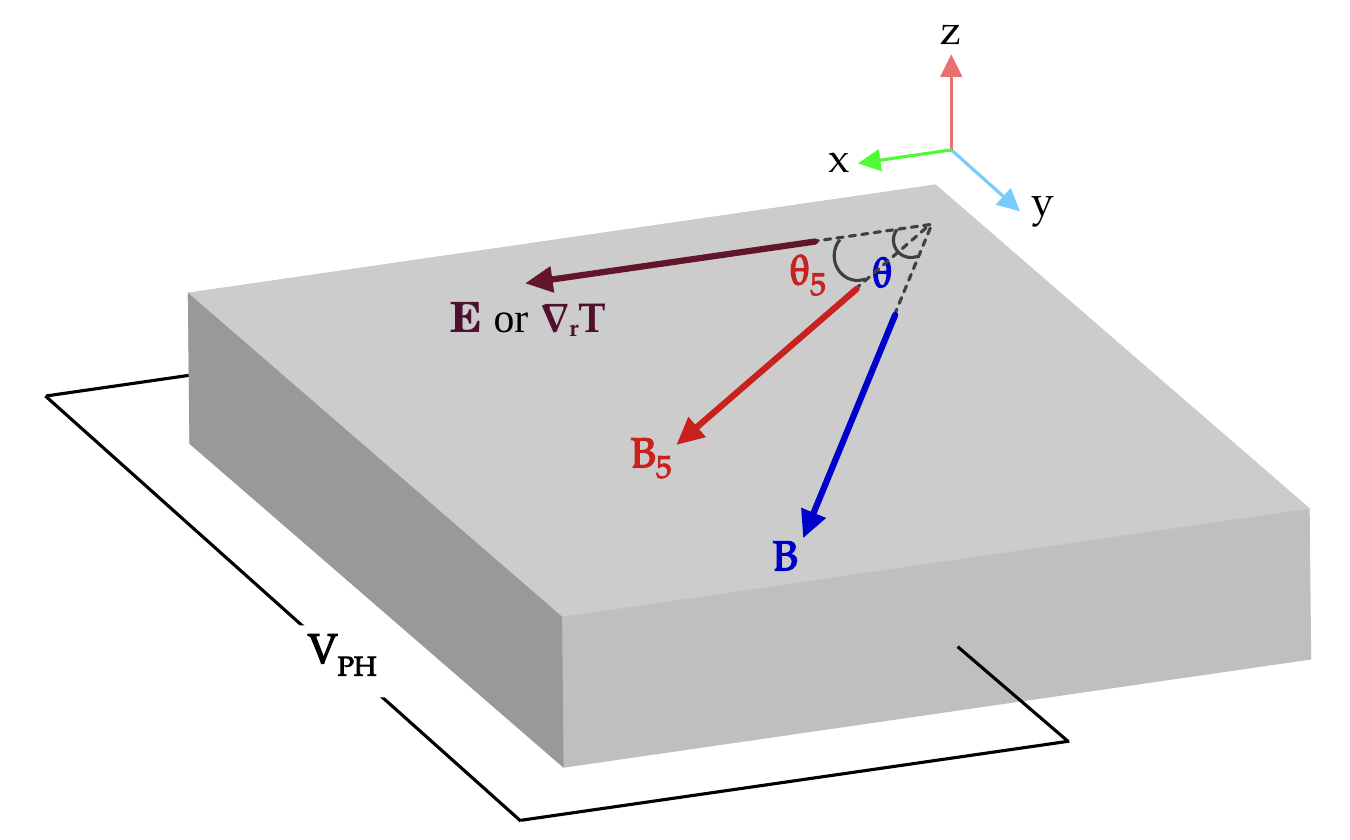}
\caption{\label{figsetup}
Schematics showing the planar Hall (planar thermal Hall) experimental set-up, where the sample is subjected to an external electric field $ E\, {\mathbf{\hat x}} $ (temperature gradient $\partial_x T\, {\mathbf{\hat x}}$). An external magnetic field $\mathbf B $ is applied such that it makes an angle $\theta $ with the existing electric field (temperature gradient). In addition, the sample is considered to be under the influence of a mechanical strain, whose effect is incorporated via an artificial chiral gauge field $\mathbf B_5 $, making an angle $\theta_5$ with the $x$-axis. The resulting planar Hall (planar thermal Hall) voltage generated along the $y$-axis is indicated by the symbol $V_{\rm PH}$.
}
\end{figure*}

The transport properties that have been widely studied in the literature include circular photogalvanic effect \cite{moore18_optical,guo23_light,kozii,ips_cpge}, circular dichroism \cite{sekh22_circular,ips_cd}, tunneling through barriers/wells \cite{mandal21_tunneling,ips-sandip,ips-sandip-sajid, ips_epj_trs}, observation of negative magnetoresistance~\cite{lv21_experimental,huang15_observation,son13_chiral,moghaddam22_observation}, intrinsic anomalous Hall effect \cite{haldane,pallab_axionic,burkov_intrinsic_hall}, planar Hall and planar thermal Hall effects~\cite{son13_chiral,burkov17_giant,nandy_2017_chiral,nandy18_Berry,nandy_thermal_hall,amit_magneto,nag21_magneto,ips-serena, *ips-mwsm-floquet}, Magnus Hall effect~\cite{papaj_magnus,das19_linear,sajid_magnus}, and magneto-optical conductivity~\cite{gusynin06_magneto,staalhammar20_magneto,yadav23_magneto}. In this paper, we will focus on the planar Hall and planar thermal Hall effects in the presence of strain.
We will compute transport coefficients involving the chiral nodes of WSMs and mWSMs. The mWSMs have hybrid anisotropic dispersions, featuring a linear dispersion along one direction which we choose to be the $z$-direction (without any loss of generality), and a quadratic/cubic dispersion in the plane perpendicular to it (i.e., along the $xy$-plane). 

Let us consider an experimental set-up with a semimetal subjected to an external electric field $ \mathbf E $ (caused by an external potential gradient) along the $x$-axis and an external magnetic field $ \mathbf B $ along the $y$-axis. Since $ \mathbf B $ is perpendicular to $ \mathbf E $, a potential difference (known as the Hall voltage) will be generated along the $z$-axis. This phenomenon is the well-known Hall effect. However, if $ \mathbf B $ is applied such that it makes an angle $ \theta $ with $ \mathbf E $, where $  \theta  \neq \pi/2$, then although the conventional Hall voltage induced from the Lorentz force is zero along the $y$-axis, transport involving a 3d semimetal node with a nonzero topological charge gives rise to a voltage difference along this direction [cf. Fig.~\ref{figsetup}]. This is known as the planar Hall effect (PHE), arising due to the chiral anomaly~\cite{son13_chiral,burkov17_giant,li_nmr17,nandy_2017_chiral,nandy18_Berry, nag21_magneto, ips-serena, *ips-mwsm-floquet}.
The associated transport coefficients related to this voltage are referred to as the longitudinal magnetoconductivity (LMC) and the planar Hall conductivity (PHC), which depend on the value of $ \theta $. In an analogous set-up, we observe the planar thermal Hall effect (PTHE) [also referred to as the planar Nernst effect (PNE)] where, instead of an external electric field, a temperature gradient $ \mathbf \nabla_{\mathbf r} T$ is applied along the $x$-axis, which then induces a potential difference along the $y$-axis due to the chiral anomaly~\cite{girish1,ips-serena, *ips-mwsm-floquet}  [cf. Fig.~\ref{figsetup}]. 
The associated transport coefficients are known as the longitudinal thermoelectric coefficient (LTEC) and transverse thermoelectric coefficient (TTEC). The behaviour of these conductivity tensors has been extensively investigated in the literature~\cite{zhang16_linear,chen16_thermoelectric,das19_linear,das20_thermal,das22_nonlinear,pal22a_berry,pal22b_berry,fu22_thermoelectric,araki20_magnetic,mizuta14_contribution}.

In a planar Hall (or thermal Hall) set-up, if a semimetal is subjected
to mechanical strain, which induces elastic deformations of the material. The elastic deformations couple to the electronic degrees of freedom (i.e., quasiparticles) in such a way that they can be modelled as pseudogauge fields in the semimetals \cite{guinea10_energy,guinea10_generating,low10_strain,landsteiner_gaguge,liu_gauge,pikulin_gauge,arjona18_rotational,onofre}. The form of these elastic gauge fields show that they couple to the quasiparticles of the Weyl nodes with opposite chiralities with opposite signs \cite{landsteiner_gaguge,liu_gauge,pikulin_gauge,ghosh20_chirality,girish2023,onofre}. Due to the chiral nature of the coupling between the emergent vector fields and the itinerant fermionic carriers, this provides an example of axial gauge fields in three dimensions. This is to be contrasted with the actual electromanegteic fields, which couple to all the nodes with the same sign. While a uniform pseudomagnetic field $ \mathbf B_5$ can be generated when a WSM/mWSM nanowire is put under torsion, a pseudoelectric field $ \mathbf E_5$ appears on dynamically stretching and compressing the crystal along an axis (which can be achieved, for example, by driving longitudinal sound waves) \cite{pikulin_gauge}. Direct evidence of the generation of such pseudoelectromagnetic fields in doped semimetals has been obtained in experiments \cite{exp_gauge}.

In this paper, we will consider co-planar scenarios with a nonzero $ \mathbf B_5$ (in the $xy$-plane), making an angle $ \theta_5$ with an actual electric field $ \mathbf E $ or a temperature gradient $\nabla_{\mathbf r} T$ applied along the $x$-axis. This is in addition to an actual co-planar magnetic field applied at angle $\theta$ with respect to the $x$-axis. The schematics of the set-up is shown in Fig.~\ref{figsetup}. We consider low magnitudes of $\mathbf B$ and $\mathbf B_5 $ such that the formation of the Landau levels can be ignored, and the magnetoelectric and magnetothermal response can be derived using the semiclassical Boltzmann formalism. The paper is organized as follows: In Sec.~\ref{sec_model}, we spell out the low-energy effective Hamiltonians for the WSMs and mWSMs. In Sec.~\ref{secph} and \ref{secphne}, we explicitly compute the conductivity tensors and discuss their behaviour in some relevant parameter regimes. Finally, we conclude with a summary and outlook in Sec.~\ref{sec_summary}. In Appendix~\ref{secboltz}, we review the semiclassical Boltzmann equations approach to derive the transport coefficients. Appendix~\ref{secstrain} is devoted to explaining the origins of the strain-induced chiral pseudomagnetic fields via explicit equations.

\begin{figure*}[t]
\subfigure[]{\includegraphics[width=0.2\linewidth]{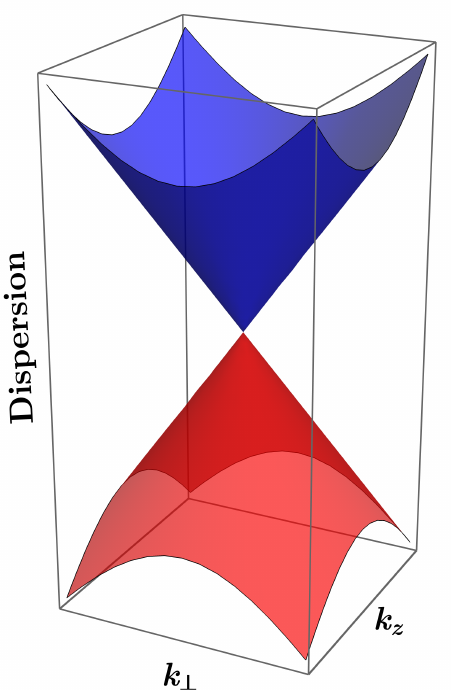}} 
\hspace{ 1.5 cm}
\subfigure[]{\includegraphics[width=0.2 \linewidth]{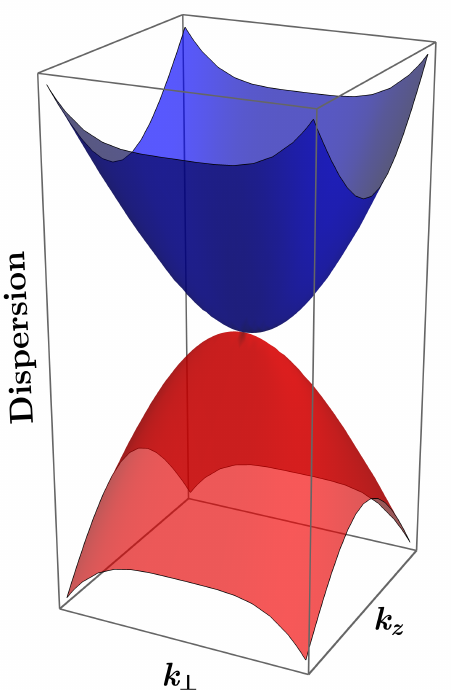}}
\hspace{ 1.5 cm}
\subfigure[]{\includegraphics[width=0.2 \linewidth]{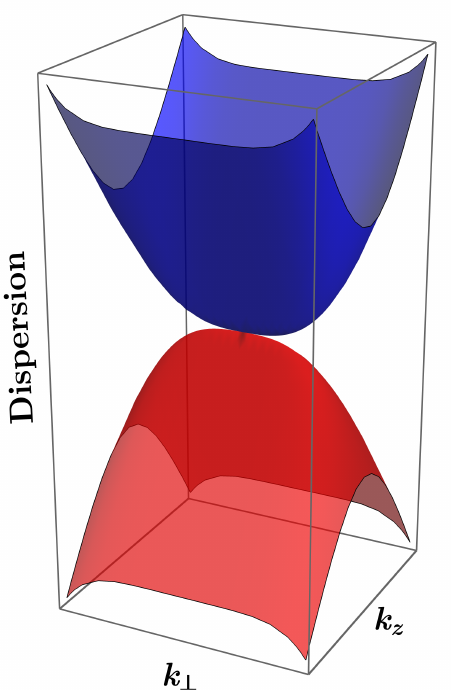}}
\caption{\label{figdis}
	Schematic dispersion of a single node in a (a) Weyl,
	(b) double-Weyl, and (c) triple-Weyl semimetal.}
\end{figure*}

\section{Model} 
\label{sec_model}

The low-energy effective Hamiltonian for a single node of WSM/mWSM can be written as \cite{liu2017predicted,bernevig,bernevig2}
\begin{align} 
\label{eqHweyl}
\mathcal{H}_\chi ( \mathbf k) & = 
\mathbf d_\chi( \mathbf k) \cdot \boldsymbol{\sigma}
+ \Delta_{\chi} \, \sigma_0\,,
\quad k_\perp=\sqrt{k_x^2 + k_y^2}\,, \quad
\phi_k=\arctan({\frac{k_y}{k_x}})\,,
\quad \alpha_J=\frac{v_\perp}{k_0^{J-1}} \,, \nn
\mathbf d_\chi( \mathbf k) &= \left \lbrace \alpha_J \, k_\perp^J \cos(J\phi_k), \,
\alpha_J \, k_\perp^J \sin(J\phi_k), \,
\chi \, v_z \, k_z \right \rbrace,
\end{align}
where $ \boldsymbol{\sigma} = \lbrace \sigma_x, \, \sigma_y, \, \sigma_z \rbrace $ is the vector operator consisting of the three Pauli matrices, $\sigma_0$ is the $2 \times 2$ identity matrix, $\chi \in \lbrace 1, -1 \rbrace $ denotes the chirality of the node, $\Delta_{\chi}$ specifies the shift in the energy at the node, and $v_z$($v_\perp$) is the Fermi velocity along the $z$-direction($xy$-plane). Here $k_0$ is a parameter with the dimension of momentum, which depends on the microscopic details of the material under consideration.
The eigenvalues of the Hamiltonian are given by
\begin{align} 
\label{eigenvalues_kperp_kz_phik}
\mathcal{E}_{\chi, s} ({ \mathbf k})= 
\Delta_{\chi} - (-1)^{s} \, \varepsilon_{\mathbf k} \,, \quad
s \in \lbrace 1,2 \rbrace ,
\quad 
\varepsilon_{\mathbf k}
= \sqrt{\alpha_J^2 \, k_\perp^{2J} + v_z^2 \, k_z^2}\,,
\end{align}
where the value $1$($2$) for $s$ represents the conduction(valence) band, as shown in Fig.~\ref{figdis}. We note that we recover the linear and isotropic nature of a WSM by setting $J=1$ and $\alpha_1= v_z$. 

The Berry curvature associated with the $s^{\rm{th}}$ band is given by  \cite{xiao10_Berry,xiao07_valley,konye21_microscopic}
\begin{align} 
\label{eqomm}
{\mathbf \Omega}_{\chi, s}( \mathbf k) & = 
i \, \langle  \nabla_{ \mathbf k}  \psi_s^\chi ({ \mathbf k})| \, \cross  \, | \nabla_{ \mathbf k}  \psi_s^\chi ({ \mathbf k})\rangle
\Rightarrow
\Omega^a_{\chi, s}( \mathbf k)  =
\frac{  (-1)^s \,  
	\epsilon^a_{\,\,\,bc}}
{4\,| \mathbf d_\chi (\mathbf k) |^3} \, 
\mathbf d_\chi (\mathbf k) \cdot
\left[   \partial_{k_b} \mathbf d_\chi (\mathbf k) \cross  \partial_{k_c} \mathbf d_\chi (\mathbf k) \right ] ,   
\end{align}
where the indices $a$, $b$, and $c$ $ \in \lbrace x, y, z \rbrace $, and are used to denote the Cartesian components of the 3d vectors and tensors. The symbol $ |  \psi_s^\chi ({ \mathbf k}) \rangle $ denotes the normalized eigenvector corresponding to the band labelled by $s$, with $ \lbrace |  \psi_1^\chi \rangle, \,  |  \psi_2^\chi \rangle \rbrace $ forming an orthonornomal set for each node. We will be working with natural units and, therefore, we set each of the constants $\hbar$, $c$, and $k_B$ to unity in all expressions in the rest of the paper (just like we have done in the starting Hamiltonian).

On evaluating the expressions in Eq.~\eqref{eqomm} using Eq.~\eqref{eqHweyl}, we get
\begin{align}
\mathbf \Omega_{\chi, s }({ \mathbf k})= 
\frac{ \chi \,(-1)^s
	J \,v_z \, \alpha_J^2 \, k_\perp^{2J-2} }
{2 \,\varepsilon^3_{\mathbf k}
} 
\left
\lbrace k_x, \, k_y, \, J\, k_z \right \rbrace .
\end{align}
The band velocity vector for the quasiparticles is given by
\begin{align}
{\boldsymbol{v}}_{\chi, s } ( \mathbf k) =
\nabla_{ \mathbf k} \,  \mathcal{E}_{\chi, s} 
({ \mathbf k}) = - \frac{ (-1)^s}
{ 
	\varepsilon_{ \mathbf k}} 
\left \lbrace J \, \alpha_J^2 \, k_\perp^{2J-2} \, k_x, \, J \, 
\alpha_J^2 \, k_\perp^{2J-2} \, k_y, \, v_z^2 \, k_z
\right \rbrace .
\end{align}
We find that $ {\boldsymbol{v}}_{\chi, s } ( \mathbf k) $ is actually independent of $\chi$.
In this paper, we will compute the transport coefficients for the case when the chemical potential cuts the conduction band. Hence, to avoid cluttering, we use the shortened notations $ \mathcal{E}_{\chi, 1}  = \mathcal{E}_{\chi} $, $\mathbf \Omega_{\chi, 1 } = \mathbf \Omega_{\chi}$, $ {\boldsymbol{v}}_{\chi, 1 } =  {\boldsymbol{v}} $, and the equilibrium Fermi-Dirac distribution function $ f^{(0)}_1 (\mathcal{E}_{\chi, 1})=  f^{(0)} (\mathcal{E}_{\chi})$.

The ranges of the values of the parameters that we will use in our computations are shown in Table~\ref{table_params}. Here we will set $\Delta_\chi = 0 $ such that the same chemical potential cuts the two nodes with opposite chiralities. In the following sections, we consider a total effective magnetic field $ \mathbf B^{\rm tot} = \mathbf B +  \chi\,\mathbf B_5 $ to be acting in the $xy$-plane, where 
\begin{align}
\mathbf B \equiv B \cos \theta \,{\mathbf{\hat x}} +  B \sin \theta \, {\mathbf{\hat y}}
\text{ and }
\mathbf B_5 \equiv B_5 \cos \theta_5 \,{\mathbf{\hat x}} +  B_5 \sin \theta_5 \, {\mathbf{\hat y}}
\end{align}
denote the parts originating from an actual magnetic field and an elastic strain field (as described in Appendix~\ref{secstrain}), respectively.

\begin{table}[t]
\begin{tabular}{|c|c|c|}
\hline
Parameter &   SI Units &   Natural Units  \\ \hline
	$v_z$ from Ref.~\cite{watzman18_dirac} & $ \sim 15\times10^{5} $ m~s$^{-1} $ 
& $\sim 0.005$  \\ \hline
	$\tau$ from Ref.~\cite{watzman18_dirac} & $ \sim 10^{-13} \, \text{s} $ & $\sim 152 $ eV$^{-1}$  \\ \hline
$ T $ from Ref.~\cite{nag21_magneto} & $ \sim 10 \, \text{K} $ & 
$ \sim 8.617 \cross 10^{-4}  $ eV \\ \hline
$ B $ and $B_5$ from Ref.~\cite{ghosh20_chirality} 
& $ 0 - 10  $ Tesla 
& $ 0 - 1950  $ eV$^{2}$
\\ \hline
$\mu$ from Refs.~\cite{Nag_2020, Nag_floquet_2020}
& $  \sim 1.6\times 10^{-21} -  1.6\times 10^{-20} $ J & $ \sim 0.01 - 0.1$ eV \\ \hline
\end{tabular}
\caption{\label{table_params}
The values of the various parameters which we have used in plotting the transport coefficients are tabulated here. In terms of natural units, we need to set $\hbar=c=k_{B}= e = 1$. In our plots, we have used $v_\perp = v_z$ (from the table entry), $\alpha_{2}=3.9 \times 10^{-5}$ eV$^{-1}$, and $\alpha_{3}=2.298 \times 10^{-6}$ eV$^{-2}$. 
For $ J=2$ and $J=3$, $v_\perp$ has been set equal to $ v_z$ for the sake of simplicity, while the isotropic dispersion for $J =1$ has $v_{\perp} = v_z$ anyway.
}
\end{table}


\section{Planar Hall set-up: Magnetoelectric transport}
\label{secph}

An electric field $\mathbf E = E \,  {\mathbf{\hat x}} $ is applied coplanar with $\mathbf B^{\rm tot}$, with zero tempearture gradient. This set-up allows us to measure the planar Hall effect.
The analysis for obtaining the expressions for the magnetoelectric coefficient tensors is explained in Appendix~\ref{secboltz}. From Eq.~\eqref{eqsigmatot}, we find that the Berry-curvature-related part of the magnetoelectric conductivity tensors, for transport via the conduction bands, are given by
\begin{align} 
\label{eqnsigma0} 
\bar \sigma_{ a b }^\chi 
=- e^2 \, \tau
\int \frac{ d^3 \mathbf k}{(2\, \pi)^3 } \, D_{\chi} 
\left [ {v}_a  + e\, B^{\rm tot}_a  \left( 
\boldsymbol{v} \cdot \mathbf \Omega_{\chi} \right)
\right ]
\left [ {v}_b  + e\,   B^{\rm tot}_b   \left( 
{\boldsymbol{v}} \cdot \mathbf \Omega_{\chi} \right)
\right ]
\, \frac{\partial  f^{(0)} } {\partial  \mathcal{E}_{\chi} } \,,
\end{align}
where
\begin{align}
\label{eqdchi}
D_{\chi}^{-1} = 1 + e \, {\mathbf B}^{\rm tot} \cdot \mathbf{\Omega }_{\chi}  \,.
\end{align} 
In order to perform the integrals, we choose to work in the limit $  e \, {\mathbf B}^{\rm tot} \cdot \mathbf{\Omega }_{\chi}  \ll 1 $ and expand $D_\chi$ up to quadratic order in $ |\mathbf B^{\rm tot} | $, such that
\begin{align}
\label{eqdchi2}
D_{\chi}  = 1 + e \, {\mathbf B}^{\rm tot} \cdot \mathbf{\Omega }_{\chi}
+ \left ( e \, {\mathbf B}^{\rm tot} \cdot \mathbf{\Omega }_{\chi} \right)^2
+\mathcal{O} \Big( \left ( e \, {\mathbf B}^{\rm tot} \cdot \mathbf{\Omega }_{\chi} \right)^3 \Big)\,.
\end{align}
The tensor component $ \bar \sigma_{ xx }^\chi $ is referred to as as the LMC, while $ \bar \sigma_{ y x }^\chi$ is known as the PHC. It is obvious that for the orientations considered here, $ \bar \sigma_{ xy }^\chi =  \bar \sigma_{ yx }^\chi $.

\begin{figure*}[t]
\centering
\includegraphics[width=0.75 \textwidth]{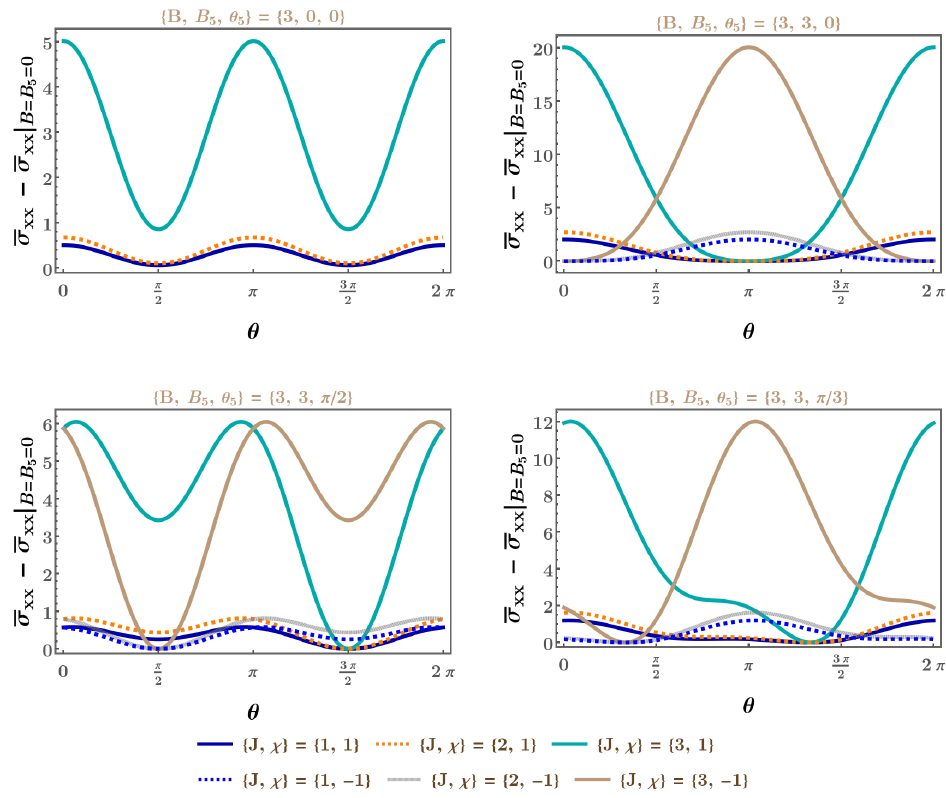}
\caption{
LMC (in units of $10^{-4} $ eV) as a function of $  \theta  $ for various values of $B$ (in units of eV$^2$), $B_5$ (in units of eV$^2$), and $\theta_5 $, with $v_z =0.005$, $\tau = 151$ eV$^{-1}$, $\mu=0.15 $ eV, and $\beta = 1160 $ eV$^{-1}$. A nontrivial dependence on the chirality $\chi$ of the node comes into play only in the presence of nonzero values of both the physical and pseudomagnetic fields. Hence, the curves shown in the first panel, where $B_5$ is set to zero, apply to both $\chi =\pm 1 $. In each of the remaining panels, since a nonzero $B_5$ is considered, the curves for $ \chi = 1$ and $\chi = -1 $ are seen to be shifted with respect to each other, as functions of the periodic variable $\theta$. The values of the maxima and minima of the curves are strongly dependent on the values of $J$.
\label{figsigxxtheta}}
\end{figure*}

\subsection{Longitudinal magnetoconductivity}

For the convenience of calculations, we first decompose $\bar \sigma_{xx}^{\chi}$ into three parts as follows:
\begin{align}
\label{eqn_sigmaXX3}
\bar \sigma_{xx}^{\chi} & = \sigma_{xx}^{\chi, {(1)}} + \sigma_{xx}^{\chi, {(2)}} 
+ \sigma_{xx}^{\chi, {(3)}}\,,\nn
\sigma_{xx}^{\chi, {(1)}} & =
\tau \, e^2  \int \frac{d^3  \mathbf k}
{( 2\, \pi)^{3}} \, D_{\chi} \, v_x^2  \,
\left ( -\frac{\partial  f^{(0)}} {\partial \mathcal{E}_{\chi}}  \right ) , \nn
\sigma_{xx}^{\chi, {(2)}} &  = \tau \, e^4 \, 
\left( B^{\rm tot}_x \right)^2  
\int \frac{d^3  \mathbf k}{( 2\, \pi)^{3}} \, D_{\chi}   
\left ( \boldsymbol v \cdot  \mathbf \Omega_{\chi} \right )^2 
\left ( -\frac{\partial  f^{(0)}} {\partial \mathcal{E}_{\chi}}  \right )  , \nn
\sigma_{xx}^{\chi, {(3)}} & = 2 \, \tau \, e^3 \, B^{\rm tot}_x  
\int \frac{d^3  \mathbf k}{( 2\, \pi)^{3}} \, D_{\chi} \, v_x
\left ( \boldsymbol v \cdot  \mathbf \Omega_{\chi} \right )
\left ( -\frac{\partial  f^{(0)}} {\partial \mathcal{E}_{\chi}}  \right )  \, .
\end{align}
To perform the integrals, we change variables as:
\begin{align}
k_\perp = \alpha_J^{-\frac{1}{J}} \, \epsilon^{\frac{1}{J}} 
\left( \sin \gamma   \right)^{\frac{1}{J}}\,,\quad
k_z = \frac{1}{v_z} \, \epsilon \,  \cos \gamma \,, \quad
k_x = k_\perp \cos \phi\,, \quad k_y = k_\perp \sin \phi\,.
\end{align}

\begin{figure}[t]
\centering
\subfigure[]{\includegraphics[width=0.95 \textwidth]{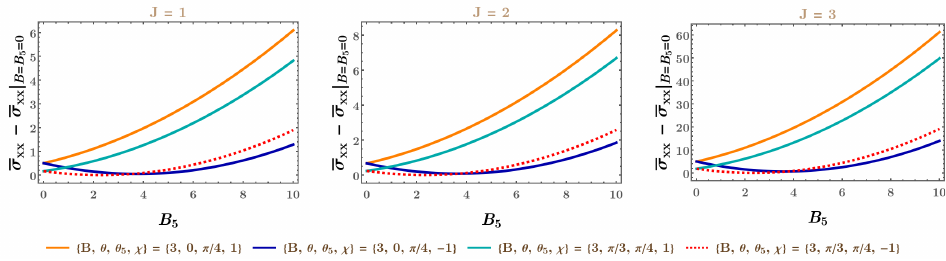}}\\
\subfigure[]{\includegraphics[width=0.95 \textwidth]{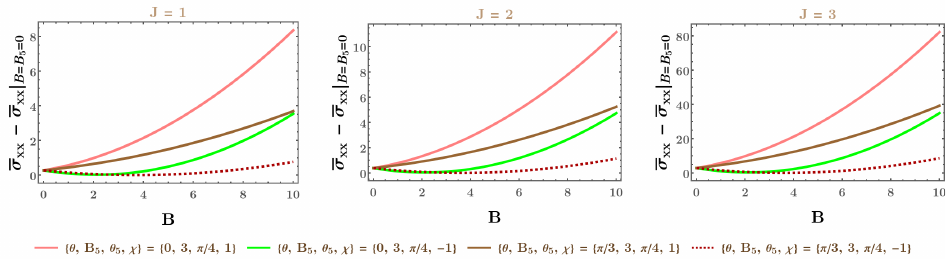}}
\caption{LMC (in units of $10^{-4} $ eV) as a function of (a) $  B_5  $ (in units of eV$^2$) with $B = 3$ eV$^2$, and (b) $B$ (in units of eV$^2$) with $B_5 = 3$ eV$^2$, for various values of $\theta $ and $\theta_5 $. We have set $v_z =0.005$, $\tau = 151$ eV$^{-1}$, $\mu=0.15 $ eV, and $\beta = 1160 $ eV$^{-1}$. The curves are parabolic in each case, with the vertex of the parabola being shifted towards right(left) for $\chi = -1$($\chi =1 $), resulting in a chirality-dependent bifurcation. However, these two curves intersect at either $B=0$ or $B_5=0$, where the chirality-dependence disappears.
\label{figsigxxB}}
\end{figure}

Using the expansion of $D_{\chi}$ [cf. Eq.~\eqref{eqdchi2}] in the low magnetic field limit, we obtain
\begin{align}
\sigma_{xx}^{\chi, {(1)}} & = \frac{\tau \, e^2 } {( 2\, \pi)^{3}} \,  \sum \limits_{n}  
\int_0^\infty dk_\perp \, \int_{-\infty}^{\infty} \, dk_z \, \int_0^{2\pi} \, d\phi \, k_{\perp}  \, v_x^2 \, 
\left [ - e  \left ( \mathbf B^{\rm tot} \cdot  \mathbf  \Omega_{\chi}  \right ) \right ]^n 
\left ( -\frac{\partial  f^{(0)}} {\partial \mathcal{E}_{\chi}}  \right ) 
\nn  & = \frac{ \tau \, e^2  }
{( 2\, \pi)^{3}} \,\frac{J} { v_z} \sum_n 
\left[-  \frac{ \chi \, e  \, v_z \, J \, 
\alpha_J^{\frac{1}{J}}} {2  } \right ]^n \,  \int_0^{2\pi} d\phi  \cos^2{\phi} \,
\left[   B  \,\cos{( \theta  - \phi)} + \chi \,B_5 \,\cos{( \theta_5 - \phi)} \right ]^n   \nn
& \hspace{ 5 cm } \times
\int_{0}^{\pi} d\gamma \left ( \sin \gamma \right )^{3 + n \,\left (2 - \frac{1}{J} \right )} 
\int_0^\infty d\epsilon \, \epsilon^{2 - n \left (1 + \frac{1}{J} \right )} 
\, \frac{\beta \, e^{\beta   \left(   \epsilon - \mu  \right)}}
{ \left [ 1 + e^{\beta \,  \left( \epsilon \, - \, \mu  \right)  } \right ]^2} \,,
\end{align}   
\begin{align} 
\sigma_{xx}^{\chi, {(2)}}   &= 
\frac{\tau \, e^4  } {( 2\, \pi)^{3}} \, 
\frac{J^3 \, v_z \, \alpha_J^{\frac{2}{J}}\, \,   \left( B_x^{\rm tot} \right)^2 }  {4}  
\sum_n   \left[- \frac{  \chi \, e \,  v_z \, J \, \alpha_J^{\frac{1}{J}}}{2} \right ] ^n 
\int_0^{2\pi} d\phi \left [  
B  \cos{( \theta  - \phi)} + \chi \, B_5  \cos{( \theta_5 - \phi)}  \right ]^n  \nn
& \hspace{ 7.3 cm } \times
\int_{0}^{\pi} d\gamma   \left ( \sin \gamma \right )^{3 -\frac{2}{J}  + n \left( 2 - \frac{1}{J} \right ) } \, 
\int_0^\infty d\epsilon \, \epsilon^{-\frac{2}{J} - n \left( 2 - \frac{1}{J} \right )} \, \frac{\beta \, e^{\beta   \left(   \epsilon - \mu  \right)}}{ \left [ 1 + e^{\beta \,  \left( \epsilon \, - \, \mu  \right)  } \right ]^2} \,,
\end{align}
\begin{align} 
\sigma_{xx}^{\chi, {(3)}} 
& =  \frac{ \tau \, e^3  }{( 2\, \pi)^{3}} \, J^2  \, 
\alpha_J^{\frac{1}{J}} \,   B_x^{\rm tot}  
\sum_n  \left[- 
\frac{  \chi \, e \,  v_z \, J \, \alpha_J^{\frac{1}{J}}}{2} \right ] ^n 
\int_0^{2\pi} d\phi 
\cos{\phi} \left [  B  \cos{( \theta  - \phi)} + \chi \, B_5  \cos{( \theta_5 - \phi)}  \right ]^n  \nn
& \hspace{6.5 cm} \times 
\int_{0}^{\pi} d\gamma  
\left  ( \sin \gamma  \right  )^{3 -\frac{1}{J} 
	+ n \left( 2 - \frac{1}{J} \right )} \, 
\int_0^\infty d\epsilon \, \epsilon^{1-\frac{1}{J} - n \left(1 + \frac{1}{J} \right )} \, 
\frac{\beta \, e^{\beta   \left(   \epsilon - \mu  \right)}}{ \left [ 1 + e^{\beta \,  \left( \epsilon \, - \, \mu  \right)  } \right ]^2}\, .
\end{align}  

For the $\epsilon$-integration, we employ the Sommerfeld expansion 
\begin{align}
\int_0^\infty d\epsilon \,\,  \epsilon^n \, \frac{\beta \, 
	e^{\beta   \, \left(   \epsilon - \mu  \right)}}
{ \left [ 1 + e^{\beta \,  \left( \epsilon  -  \mu  \right)  } \right ]^2} =  
\mu^n \left [ 1 + \frac{\pi^2 \, n \, (n - 1)}{6 \, \beta^2 \, \mu^2}
+ \ldots   \right  ] \,,
\end{align}
which is applicable under the condition $\beta \mu \gg 1$. Hence, we choose appropriate values of $\beta$ and $\mu$ such that  we are using this approximation in the correct regime.
Keeping terms upto quadratic order in the components of $  \mathbf B^{\rm tot} $, the final expressions are given by
\begin{align}
\sigma_{xx}^{\chi, {(1)}} & = 
\frac{  \tau \, e^2 \,  J \,}{6\, \pi^2 \, v_z} \mu^2 
\left(1 + \frac{\pi ^2} {3 \, \beta^2 \, \mu^2} \right) 
+ 
\frac{ \tau \, e^4 v_z \, J^3 \, \alpha_J^{\frac{2}{J}} \, \mu^{-\frac{2}{J}}
}
{128 \, \pi^{\frac{3}{2}} }   
\left[ 3 \,  \left ( B_x^{\rm tot} \right ) ^2 
+   \left ( B_y^{\rm tot} \right )^2 \right] 
\frac{ \Gamma (4-\frac{1}{J})}
{\Gamma (\frac{9}{2}-\frac{1}{J})} 
\left [ 1 +  \frac{ \pi^2 \, (J + 2)}{3 \, \beta^2 \, \mu^2 \, J^2} \right ] ,\nn
\sigma_{xx}^{\chi, {(2)}} & = 
\frac{\tau \, e^4 \, v_z \, J^3 \alpha_J^{\frac{2}{J} }
	\,  \mu^{-\frac{2}{J}} }
{16 \, 
	\pi^{\frac{3}{2}}} \,  \left ( B_x^{\rm tot} \right )^2  \,\, \frac{ \Gamma(2-\frac{1}{J})}{\Gamma (\frac{5}{2}-\frac{1}{J})} 
\Big [ 1+ \frac{\pi^2 \, (J + 2)}{3 \, \beta^2 \, \mu^2 \, J^2}  \Big ] ,\nn
\sigma_{xx}^{\chi, {(3)}} & = - 
\frac{\tau \, e^4 \, v_z \, J^3 \,
	\alpha_J^{\frac{2}{J}} \,  \mu^{-\frac{2}{J}} }
{16 \, \pi^{\frac{3}{2}}} \,  \left ( B_x^{\rm tot} \right )^2  \,\, \frac{ \Gamma(3-\frac{1}{J})}{\Gamma (\frac{7}{2}-\frac{1}{J})} 
\left[  1 + \frac{\pi^2 \, (J + 2)}{3 \, \beta^2 \, \mu^2 \, J^2}  \right ].
\end{align}

Adding all the three parts, we have
\begin{align} 
\label{eqn:sigmaxxfinal}
\bar  \sigma_{xx}^{\chi} & =   \frac{  \tau \, e^2 \, J \, \mu^2  }
{6\, \pi^2 \, v_z} 
\left[ 1 + \frac{\pi ^2} {3 \, \beta^2 \, \mu^2} \right ]
+  
\frac{ \tau \, e^4 \, v_z \, J \, \alpha_J^{\frac{2}{J}} \, \mu^{- \frac{2}{J}}}
{128 \, \pi^{\frac{3}{2}} } 
\left [   N_{xx1}   \left ( B_x^{\rm tot} \right )^2 
+ N_{xx2} \left ( B_y^{\rm tot} \right )^2 \right]
\frac{ \Gamma (2-\frac{1}{J})}{\Gamma (\frac{9}{2}-\frac{1}{J})} 
\left[ 1 +  \frac{ \pi^2 \,
	(J + 2)}{3 \, \beta^2 \, \mu^2 \, J^2} \right ], \nn
N_{xx1}  & = 32 \,J^2 -19 \, J +3\,, \quad
N_{xx2}  = 6\, J^2 - 5 \, J + 1 \,.
\end{align}
Clearly, $N_{xx1}$ and $N_{xx2}$ are both positive. The first term in $\bar \sigma_{xx}^{\chi} $ is independent of the magnetic field, varies linearly with $J$, and has a nonzero value even at zero temperature. This $ \mathbf B^{\rm tot} $-independent part is usually referred to as the Drude contribution. Since $ \mathbf B^{\rm tot} = \mathbf B +  \chi\,\mathbf B_5 $, it is clear from Eq.~\eqref{eqn:sigmaxxfinal} that the chirality-independent part of $\bar  \sigma_{xx}^{\chi}$ is proportional to $ (B^2 + B_5^2 ) $ modulo a $\mathbf B_5 $-independent part, while the part proportional to $\chi $ varies as linear-in-$B_5$.

The behaviour of the LMC, for some representative parameters, is shown in Figs.~\ref{figsigxxtheta} and \ref{figsigxxB}.
Fig.~\ref{figsigxxtheta} illustrates the behaviour of LMC as a function of $  \theta  $ for various values of $B$, $B_5$, and $\theta_5 $. The dependence on the chirality of the node comes into play only when both $  B$ and $  B_5 $ take nonzero values. Hence, in the first panel, where $B_5$ is set to zero, the curves for both chiralities coincide. Each of the remaining panels involves nonzero $B_5$ values, and the curves for $ \chi = 1$ and $\chi = -1 $ are seen to be shifted with respect to each other. The values of the maxima and minima of the curves are strongly dependent on the values of $J$, as expected from the expressions in Eq.~\eqref{eqn:sigmaxxfinal}. Fig.~\ref{figsigxxB} demonstrates the dependence of the LMC on $B_5$ and $B $, which captures a generic parabolic behaviour (with respect to each of the variables) shown in Eq.~\eqref{eqn:sigmaxxfinal}. The vertex of the parabola in each case shifts towards right for $\chi = -1$ and towards left for $\chi =1 $, which results in a bifurcation of the curves representing them. These two curves of course intersect when either $B$ or $B_5$ goes to zero, because this condition makes the chirality-dependence disappear.

\begin{figure}[t]
\centering
\subfigure[]{\includegraphics[width=0.45 \textwidth]{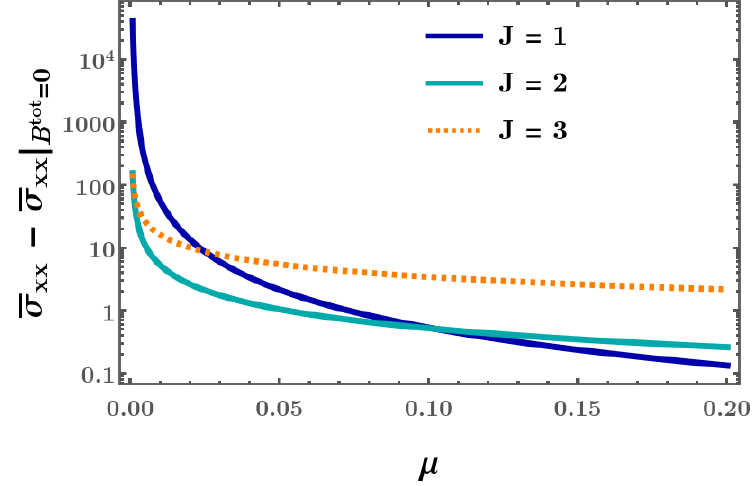}}\quad
\subfigure[]{\includegraphics[width=0.45 \textwidth]{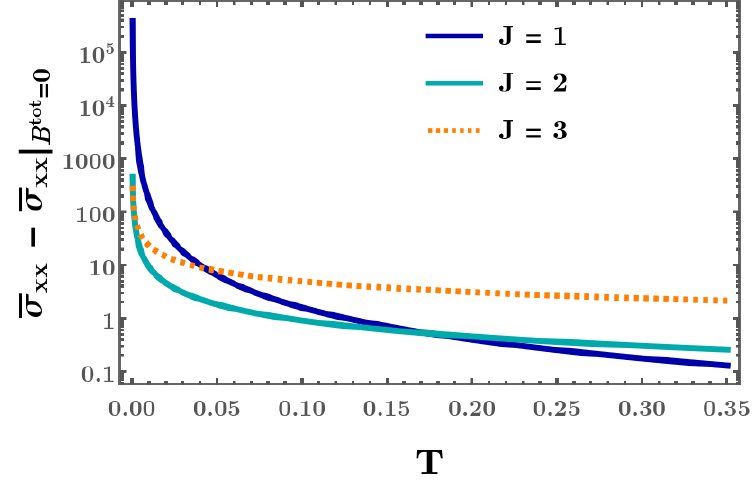}}
\caption{
LMC (in units of $10^{-4} $ eV) as a function of (a) $  \mu  $ (in units of eV) with $ T = {8.617} \times 10^{-4} $ eV, and (b) $ T $ (in units of eV) with $ \mu = 0.1 $ eV, for $v_z =0.005$, $\tau = 151$ eV$^{-1}$, $ B = 3 $ eV, and $\theta = \pi/3 $. For simplicty, we have set $B_5 = 0$.
\label{figcross}}
\end{figure}

The relative magnitudes of the curves are strongly dependent on the values of $J$, but in a complicated way. The expressions for the conductivity tensor contain factors like $\left( \alpha_J/ \mu  \right)^{2/J}$, and hence the overall response for a given $J$ depends crucially on the value of $\mu$. This is illustrated via Fig.~\ref{figcross}, where we have set $B_5 = 0 $ for simplicity.
Of course, extremely low values of $\mu$ (e.g., $ \mu \lesssim 0.001$ for $ T \simeq {8.617} \times 10^{-4} $ eV) are not admissible because the Sommerfeld expansion is applicable only for $\beta \,\mu \gg 1$.

\begin{figure}[t]
\centering
\includegraphics[width=0.75 \textwidth]{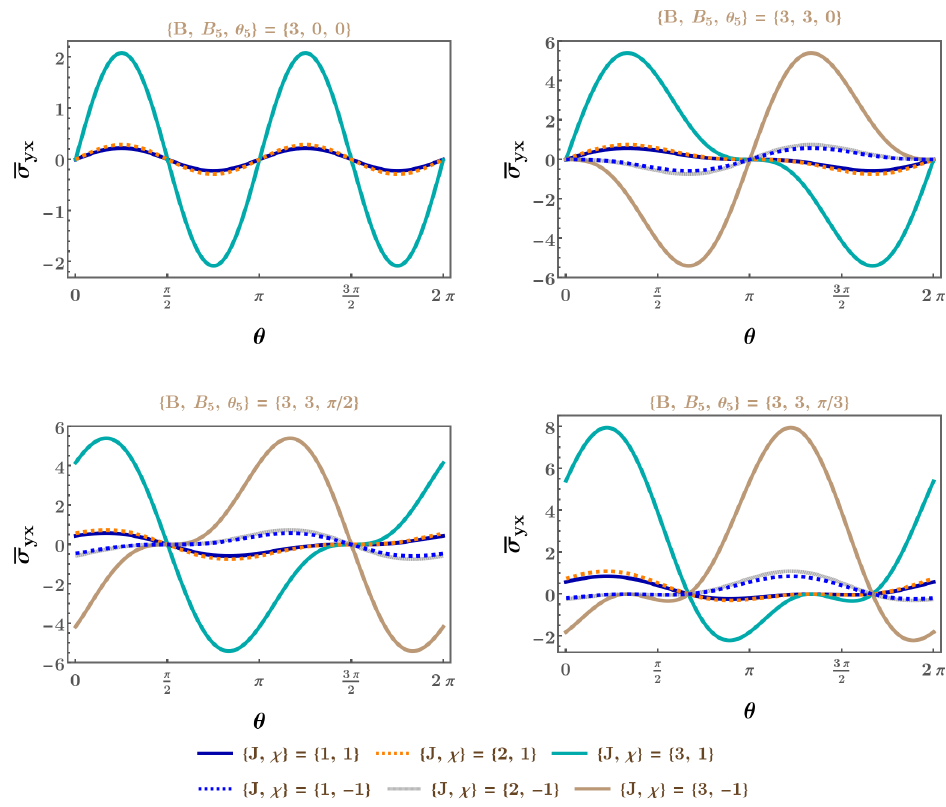}
\caption{PHC (in units of $10^{-4} $ eV) as a function of $  \theta  $ for various values of $B$ (in units of eV$^2$), $B_5$ (in units of eV$^2$), and $\theta_5 $, with $v_z =0.005$, $\tau = 151$ eV$^{-1}$, $\mu=0.15 $ eV, and $\beta = 1160 $ eV$^{-1}$. A nontrivial dependence on the chirality $\chi$ of the node comes into play only in the presence of nonzero values of both the physical and pseudomagnetic fields. Hence, the curves shown in the first panel, where $B_5$ is set to zero, apply to both $\chi =\pm 1 $. In each of the remaining panels, since a nonzero $B_5$ is considered, the curves for $ \chi = 1$ and $\chi = -1 $ are seen to be shifted with respect to each other, as functions of the periodic variable $\theta$. The values of the maxima and minima of the curves are strongly dependent on the values of $J$.
\label{figsigxytheta}}
\end{figure}


\subsection{Planar Hall conductivity}

In order to obtain the analytical expressions for the PHC, similar to the treatment of $\bar \sigma_{xx}^\chi$, we first decompose 
$ \sigma_{y x }^{\chi} $ into four parts as follows:
\begin{align}
\bar \sigma_{y x }^{\chi} &= \sigma_{y x }^{\chi, {(1)}} + \sigma_{y x }^{\chi, {(2)}} 
+ \sigma_{y x }^{\chi, {(3)}} + \sigma_{y x }^{\chi, {(4)}}
\,, \nn
\sigma_{y x }^{\chi, {(1)}} & = 
\tau \, e^2  \int \frac{d^3  \mathbf k}
{( 2\, \pi)^{3}} \, D_\chi \,
v_x \, v_y \, \left ( -\frac{\partial  f^{(0)}} {\partial \mathcal{E}_{\chi}}  \right )  ,\quad
\sigma_{y x }^{\chi, {(2)}} = \tau \, e^4 \,  B^{\rm tot}_{x} \, B^{\rm tot}_{y} 
\int \frac{d^3  \mathbf k}{( 2\, \pi)^{3}} \, 
D_\chi  \left  ( \boldsymbol v \cdot  \mathbf \Omega_\chi \right )^2
\left ( -\frac{\partial  f^{(0)}} {\partial \mathcal{E}_{\chi}}  \right )  , \nn
\sigma_{y x }^{\chi, {(3)}} & =  \tau \, e^3 \,B^{\rm tot}_{y} 
\int \frac{d^3  \mathbf k}{( 2\, \pi)^{3}} \, D_\chi \, v_x \,
\left  ( \boldsymbol v \cdot  \mathbf \Omega_\chi \right ) \,
\left ( -\frac{\partial  f^{(0)}} {\partial \mathcal{E}_{\chi}}  \right ) ,  
\quad
\sigma_{y x }^{\chi, {(4)}}
=  \tau \, e^3 \, B^{\rm tot}_x
\int \frac{d^3  \mathbf k}{( 2\, \pi)^{3}} \, D_\chi \, 
v_y \,   \left  ( \boldsymbol v \cdot  \mathbf \Omega_\chi \right ) 
\left ( -\frac{\partial  f^{(0)}} {\partial \mathcal{E}_{\chi}}  \right ) .
\end{align}
Adopting the same strategy as in the LMC case, we get the expressions
\begin{align}
\sigma_{y x }^{\chi, {(1)}} & =  \frac{ \tau \, e^4 \,v_z \, J^3 \, 
	\alpha_J^{\frac{2}{J}} \, \mu^{-\frac{2}{J}}}
{64 \, \pi^{\frac{3}{2}} } \,    
B^{\rm tot}_{x} \,  B^{\rm tot}_{y}  \, \frac{ \Gamma (4-\frac{1}{J})}
{\Gamma (\frac{9}{2}-\frac{1}{J})} 
\left [ 1 +  \frac{ \pi^2 \, (J + 2)}{3 \, \beta^2 \, \mu^2 \, J^2}\right ]  , \nn
\sigma_{y x }^{\chi, {(2)}} &  = \frac{ B^{\rm tot}_y }
{  B^{\rm tot}_x} \, \sigma_{xx}^{\chi, {(2)}} , \,\,
\sigma_{y x }^{\chi, {(3)}} =  \sigma_{y x }^{\chi, {(4)}}
=  \frac{  B^{\rm tot}_y } { 2\, B^{\rm tot}_x } \,
\sigma_{x x }^{\chi, {(3)}}  .
\end{align}
Adding all the four parts, the final form is obtained as
\begin{align} 
\label{eqn:sigmaxyfinal}
\bar  \sigma_{y x }^{\chi} =  \frac{ \tau \, e^4 \,v_z \, J \, 
	\alpha_J^{\frac{2}{J}} \, \mu^{-\frac{2}{J}}   }
{64 \, \pi^{\frac{3}{2}} } \, 
N_{y x }  \,B^{\rm tot}_{x} \,  B^{\rm tot}_{y}  \, 
\frac{ \Gamma (2-\frac{1}{J})}{\Gamma (\frac{9}{2}-\frac{1}{J})}  
\left [ 1 +  \frac{ \pi^2 \,(J + 2)}
{3 \, \beta^2 \, \mu^2 \, J^2} \right ] , \quad
N_{y x } =   13 J^2 - 7 J +1 \,. 
\end{align}
Here we can see that PHC is proportional to $  B^{\rm tot}_{x} \, B^{\rm tot}_{y}$. Therefore, for $ \sin \theta =0 $($ \cos \theta =0  $), we get a linear-in-$B$ dependence (modulo a $B$-independent shift) in the presence of a $\mathbf B_5 $ with a nonzero $y$-component($x$-component). This is to be contrasted with the situation in the absence of a pseudomagnetic field, because then the response is zero when $\mathbf B $ is oriented parallel, anti-parallel,  or perpendicular to $\mathbf E$.


\begin{figure}
\centering
\subfigure[]{\includegraphics[width=0.95 \textwidth]{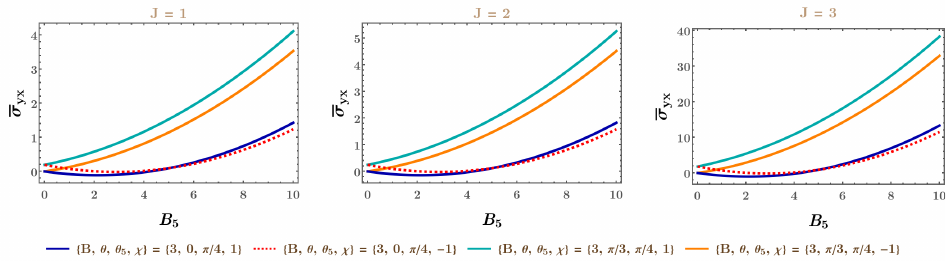}} \\
\subfigure[]{\includegraphics[width=0.95 \textwidth]{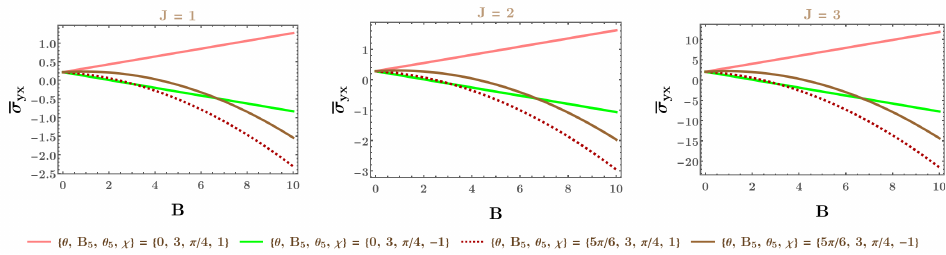}}
\caption{
PHC (in units of $10^{-4} $ eV) as a function of (a) $  B_5  $ (in units of eV$^2$) with $B = 3$ eV$^2$, and (b) $B$ (in units of eV$^2$) with $B_5 = 3$ eV$^2$, for various values of $\theta $ and $\theta_5 $. We have set $v_z =0.005$, $\tau = 151$ eV$^{-1}$, $\mu=0.15 $ eV, and $\beta = 1160 $ eV$^{-1}$. The curves are parabolic in each case, with the vertex of the parabola being shifted towards right(left) for $\chi = -1$($\chi =1 $), resulting in a chirality-dependent bifurcation. However, for special cases, the dependence is linear. For $ B \cos \theta = 0 $($ B \sin \theta = 0 $), the response is linear-in-$B$ as long as $ B_5 \cos \theta_5 $($ B_5 \sin \theta_5 $) is nonzero. Analogously, for $ B_5 \cos \theta_5 = 0 $($ B_5 \sin \theta_5 = 0 $), the response is linear-in-$B_5$ as long as $ B \cos \theta $($ B \sin \theta $) is nonzero, and is directly proportional to $\chi $. The pink and green lines of subfigure (b) demonstrate such a special case when $\theta =0 $.
\label{figsigxyB}}
\end{figure}

The behaviour of the PHC, for some representative parameters, is shown in Figs.~\ref{figsigxytheta} and \ref{figsigxyB}. Fig.~\ref{figsigxytheta} illustrates the behaviour of the PHC as a function of $  \theta  $ for various values of $B$, $B_5$, and $\theta_5 $. The dependence on the chirality of the node comes into play only when both $B$ and $  B_5 $ take nonzero values. Hence, in the first panel, where $B_5$ is set to zero, the curves for both chiralities coincide. Each of the remaining panels involves nonzero $B_5$ values, and the curves for $ \chi = 1$ and $\chi = -1 $ are seen to be shifted with respect to each other. The values of the maxima and minima of the curves are strongly dependent on the values of $J$, as expected from the expressions in Eq.~\eqref{eqn:sigmaxyfinal}. Fig.~\ref{figsigxyB} represents the dependence of the PHC on $B_5$ and $B $, which captures the parabolic behaviour (with respect to each of the variables), except for special cases. For $ B \cos \theta = 0 $($ B \sin \theta = 0 $), the response is linear-in-$B$ as long as $ B_5 \cos \theta_5 $($ B_5 \sin \theta_5 $) is nonzero. Analogously, for $ B_5 \cos \theta_5 = 0 $($ B_5 \sin \theta_5 = 0 $), the response is linear-in-$B_5$ as long as $ B \cos \theta $($ B \sin \theta $) is nonzero, and is directly proportional to $\chi $. The pink and green lines of Fig.~\ref{figsigxyB}(b) demonstrate such a special case. For the generic cases of parabolic curves, the vertex of the parabola in each case shifts towards right for $\chi = -1$ and towards left for $\chi =1 $. The two curves diverge from each other from the point where either $B$ or $B_5$ goes to zero. Just like the case of LMC, the relative magnitudes of the curves are strongly dependent on the values of $J$ via a complex functional dependence involving $\mu$ as well.

\begin{figure*}[t]
\centering
\includegraphics[width=0.75 \textwidth]{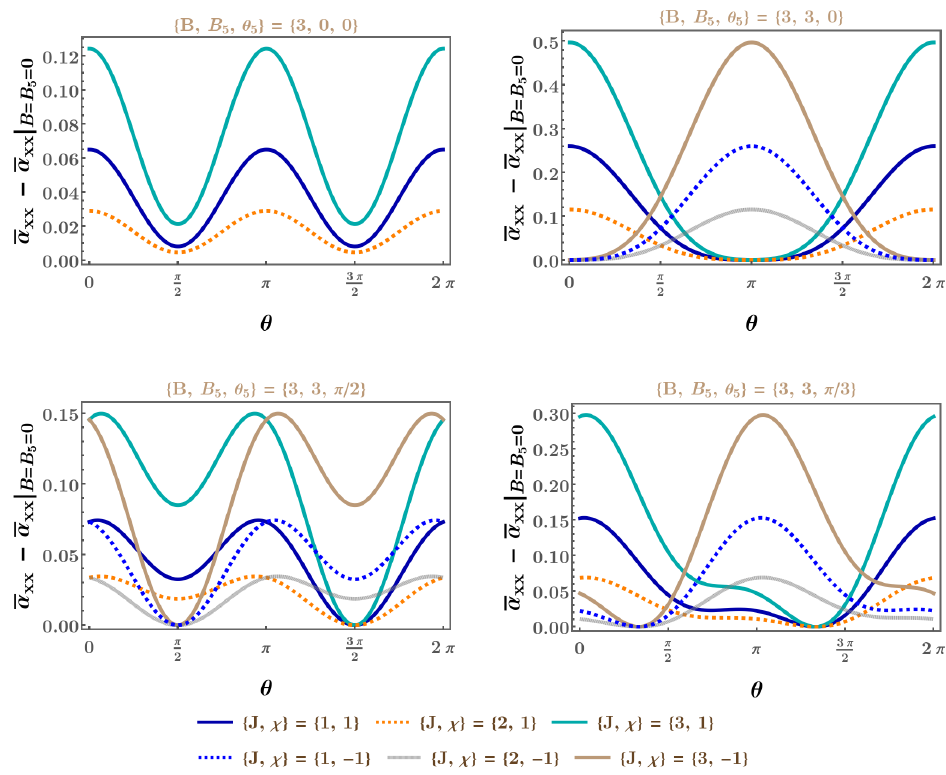}
\caption{
LTEC (in units of $10^{-4} $ eV) as a function of $  \theta  $ for various values of $B$ (in units of eV$^2$), $B_5$ (in units of eV$^2$), and $\theta_5 $, with $v_z =0.005$, $\tau = 151$ eV$^{-1}$, $\mu=0.1 $ eV, and $\beta = 1160 $ eV$^{-1}$. A nontrivial dependence on the chirality $\chi$ of the node comes into play only in the presence of nonzero values of both the physical and pseudomagnetic fields. Hence, the curves shown in the first panel, where $B_5$ is set to zero, apply to both $\chi =\pm 1 $. In each of the remaining panels, since a nonzero $B_5$ is considered, the curves for $ \chi = 1$ and $\chi = -1 $ are seen to be shifted with respect to each other, as functions of the periodic variable $\theta$. The values of the maxima and minima of the curves are strongly dependent on the values of $J$.
\label{figalphaxxtheta}}
\end{figure*}


\section{Planar thermal Hall set-up: Magnetothermal transport}   
\label{secphne}

A nonzero temperature gradient $ \nabla_{\mathbf r } T = \partial_x T \,  {\mathbf{\hat x}} $ is applied, such that it is coplanar with $\mathbf B^{\rm tot}$. There is no applied voltage, i.e., $\mathbf E = 0 $. This set-up allows us to measure the planar thermal Hall effect.
The analysis for obtaining the expressions for the magnetothermal conductivity tensors is explained in Appendix~\ref{secboltz}. From Eq.~\eqref{eqalphatot}, we find that the Berry-curvature-related part of the thermoelectric coefficient, applicable for the conduction bands, is given by
\begin{align} 
\label{eqnsigma} 
\bar \alpha_{ a b }^\chi 
= e \, \tau
\int \frac{ d^3 \mathbf k}{(2\, \pi)^3 } \, D_{\chi} 
\left [ { v}_a  + e\, B^{\rm tot}_a  \left( 
\boldsymbol{v} \cdot \mathbf \Omega_{\chi} \right)
\right ]
\left [ { v }_b  + e\,  B^{\rm tot}_b   \left( 
{\boldsymbol{v}} \cdot \mathbf \Omega_{\chi} \right)
\right ] \,
\frac{  \mathcal E_\chi -\mu }  {T} 
\, \frac{\partial  f^{(0)} } {\partial  \mathcal{E}_{\chi} } \,,
\end{align}
with $D_{\chi}$ defined in Eq.~\eqref{eqdchi}.
The tensor component $ \bar \alpha_{ xx }^\chi $ is referred to as as the LTEC, while $ \bar \sigma_{ y x }^\chi$ is known as the TTEC. Again, due to the symmetry of the dispersions of the WSMs/mWSMs in the $xy$-plane, $ \bar \alpha_{ xy }^\chi =  \bar \alpha_{ yx }^\chi $.

\begin{figure}
\centering
\subfigure[]{\includegraphics[width=0.95 \textwidth]{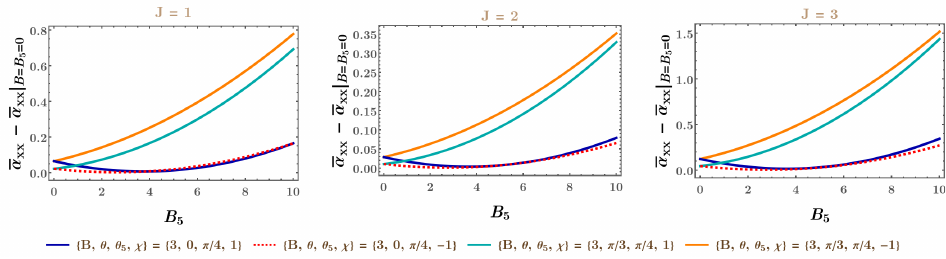}} \\
\subfigure[]{\includegraphics[width=0.95 \textwidth]{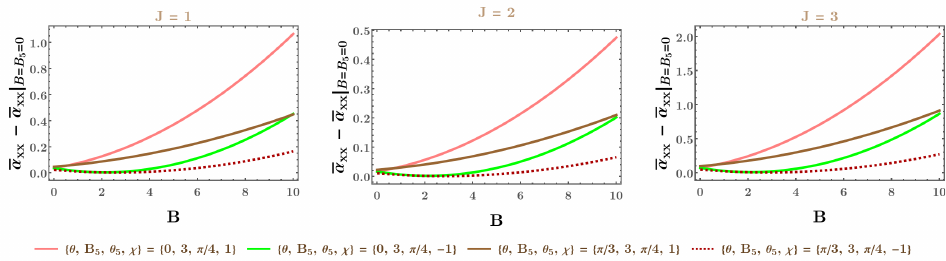}}
\caption{
	LTEC (in units of $10^{-4} $ eV) as a function of (a) $  B_5  $ (in units of eV$^2$) with $B = 3$ eV$^2$, and (b) $B$ (in units of eV$^2$) with $B_5 = 3$ eV$^2$, for various values of $\theta $ and $\theta_5 $. We have set $v_z =0.005$, $\tau = 151$ eV$^{-1}$, $\mu=0.1 $ eV, and $\beta = 1160 $ eV$^{-1}$. The curves are parabolic in each case, with the vertex of the parabola being shifted towards right(left) for $\chi = -1$($\chi =1 $), resulting in a chirality-dependent bifurcation. However, these two curves intersect at either $B = 0$ or $B_5 =0 $, where the chirality-dependence disappears.
	\label{figalphaxxB}}
\end{figure}

\subsection{Longitudinal thermoelectric coefficient}

For the ease of calculations, we decompose $\bar \alpha_{xx}$ into three parts as follows:
\begin{align}
\label{eqalphaxx}
\bar \alpha_{xx}^{\chi} & = \alpha_{xx}^{\chi, {(1)}} + \alpha_{xx}^{\chi, {(2)}} 
+ \alpha_{xx}^{\chi, {(3)}}\,,\nn
\alpha_{xx}^{\chi, {(1)}} & = 
\tau \, e \int \frac{d^3  \mathbf k}{( 2\, \pi)^{3}} \, D_{\chi} \, v_x^2 \, \,
\frac{  \mathcal E_\chi -\mu }  {T}\, \frac{\partial  f^{(0)}} {\partial \mathcal{E}_{\chi}} \, , \nn
\alpha_{xx}^{\chi, {(2)}} &  = \tau \, e^3 \, 
\left(  B^{\rm tot}_x  \right)^2
\int \frac{d^3  \mathbf k}{( 2\, \pi)^{3}} \, D_{\chi}   
\left ( \boldsymbol v \cdot  \mathbf \Omega_{\chi} \right )^2 \,
\frac{  \mathcal E_\chi -\mu }  {T}\,
\frac{\partial  f^{(0)}} {\partial \mathcal{E}_{\chi}} \, , \nn
\alpha_{xx}^{\chi, {(3)}} & = 
2 \, \tau \, e^2 \, B^{\rm tot}_x
\int \frac{d^3  \mathbf k}{( 2\, \pi)^{3}} \, D_{\chi} \, v_x
\left ( \boldsymbol v \cdot  \mathbf \Omega_{\chi} \right )\,
\frac{  \mathcal E_\chi -\mu }  {T}\,
\frac{\partial  f^{(0)}} {\partial \mathcal{E}_{\chi}}    \, .
\end{align}
Evaluating the integrals, we get
\begin{align}
\label{eqalphaxx2}
\alpha_{xx}^{\chi, {(1)}} &= 
-\frac{  \tau \, e \,  J \, \mu} {9 \, v_z \, \beta} 
+ \frac{ \tau \, e^3 \,v_z \, J^2 \, \alpha_J^{\frac{2}{J}  } 
\, \sqrt \pi   \, \mu^{-1-\frac{2}{J}} \,\zeta
}
{192 \, \beta } 
\left [ 3 \,  \left ( B_x^{\rm tot} \right  )^2 +   
\left ( B_y^{\rm tot} \right )^2 \right]
\,  \frac{ \Gamma (4-\frac{1}{J})}{\Gamma (\frac{9}{2}-\frac{1}{J})} \, ,\nn
\alpha_{xx}^{\chi, {(2)}} &=
\frac{ \tau \, e^3 \,v_z \, J^2 \, \alpha_J^{\frac{2}{J}} \,  {\sqrt \pi}
\,	\mu^{ -1 - \frac{2}{J}} \,\zeta    }
{24 \, \beta } 
\left ( B_x^{\rm tot} \right )^2 \, 
\frac{ \Gamma (2 - \frac{1}{J})}
{\Gamma (\frac{5}{2} - \frac{1}{J})} \, , \quad
\alpha_{xx}^{\chi, {(3)}}  = -  
\frac{ \tau \, e^3 \,v_z \, J^2 \, \alpha_J^{\frac{2}{J}} \, 
{\sqrt \pi}  \, \mu^{ -1 - \frac{2}{J}} \,\zeta } 
{24 \, \beta } \,  \left ( B_x^{\rm tot} \right )^2 \, 
\frac{ \Gamma (3 - \frac{1}{J})}{\Gamma (\frac{7}{2} - \frac{1}{J})} \,,
\end{align}
where
\begin{align}
\zeta = 1+\frac{\pi^2 \,(J+1)\,(J+2)}
{3 \, \beta^2\, \mu^2 \,J^2} \,.
\end{align}

The final expression for the LTEC turns out to be
\begin{align}
\bar \alpha_{xx}^{\chi} =
-\frac{  \tau \, e \, J \, \mu}{9 \, v_z \, \beta} 
+ \frac{ \tau \, e^3 \,J^2\,
v_z \, \alpha_J^{\frac{2}{J}} \,  {\sqrt \pi}\, \mu^{-1-\frac{2}{J}}
\,\zeta
}
{192 \,\beta }
\left [ N_{xx1} \,  \left ( B_x^{\rm tot} \right )^2 + N_{xx2}  
\left ( B_y^{\rm tot} \right )^2 \right ] \, \frac{ \Gamma (2 -\frac{1}{J})}
{\Gamma (\frac{9}{2}-\frac{1}{J})}   \,.
\end{align}
Comparing with Eq.~\eqref{eqn:sigmaxxfinal}, we observe that $
\partial_\mu \bar \sigma_{xx}^{\chi} = - \frac  {3\, e\, \beta }  {\pi^2} \,\bar \alpha_{xx}^{\chi}
+ \mathcal{O} (\beta^{-2}) $. Hence, the Mott relation $ L_{ab}^{12}  =  - \frac{\pi^2} {3\, e\, \beta } \,\partial_\mu L_{ab}^{11} $ [where the $L_{ab}$'s have been defined in Eq.~\eqref{eqcur1}], which holds in the $\beta \rightarrow \infty $ limit, is satisfied \cite{xiao06_berry}.

The behaviour of the LTEC, for some representative parameters, is shown in Figs.~\ref{figalphaxxtheta} and \ref{figalphaxxB}. Fig.~\ref{figalphaxxtheta} illustrates the behaviour of the LTEC as a function of $  \theta  $ for various values of $B$, $B_5$, and $\theta_5 $. Fig.~\ref{figalphaxxB} illustrates the dependence of the LTEC on $B_5$ and $B $. The overall features are similar to those observed for the LMC and, hence, in order to capture a slightly different parameter regime, we use a somewhat lower value of $\mu $ than that used for the LMC curves.

\begin{figure}
\centering
\includegraphics[width=0.75 \textwidth]{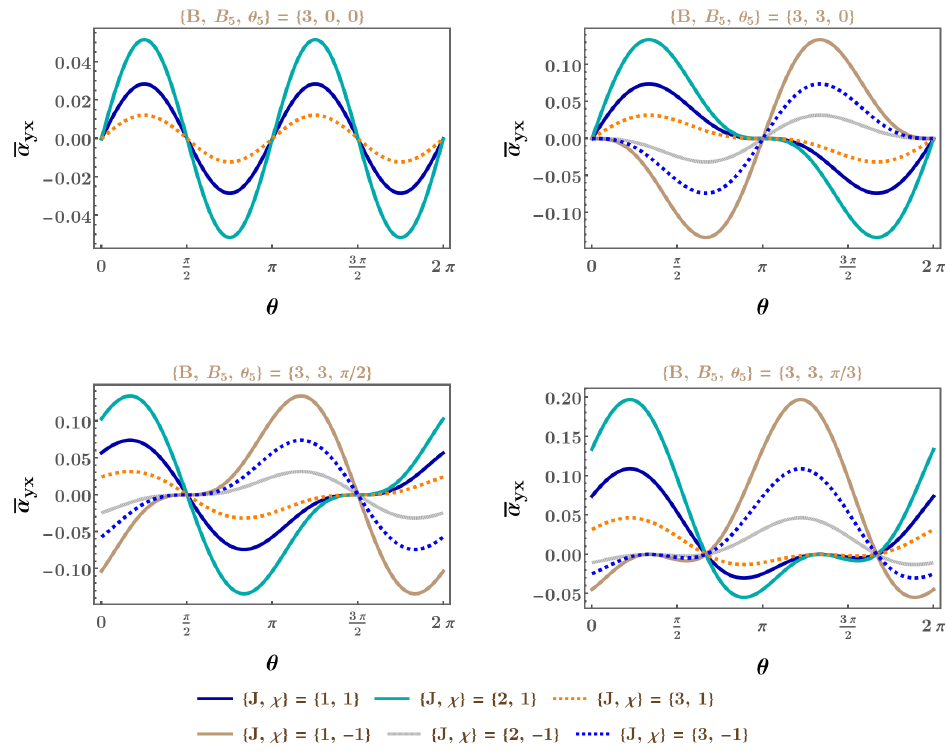}
\caption{
	TTEC (in units of $10^{-4} $ eV) as a function of $  \theta  $ for various values of $B$ (in units of eV$^2$), $B_5$ (in units of eV$^2$), and $\theta_5 $, with $v_z =0.005$, $\tau = 151$ eV$^{-1}$, $\mu=0.1 $ eV, and $\beta = 1160 $ eV$^{-1}$. A nontrivial dependence on the chirality $\chi$ of the node comes into play only in the presence of nonzero values of both the physical and pseudomagnetic fields. Hence, the curves shown in the first panel, where $B_5$ is set to zero, apply to both $\chi =\pm 1 $. In each of the remaining panels, since a nonzero $B_5$ is considered, the curves for $ \chi = 1$ and $\chi = -1 $ are seen to be shifted with respect to each other, as functions of the periodic variable $\theta$. The values of the maxima and minima of the curves are strongly dependent on the values of $J$.
	\label{figalphaxytheta}}
\end{figure}

\begin{figure}
\centering
\subfigure[]{\includegraphics[width=0.95 \textwidth]{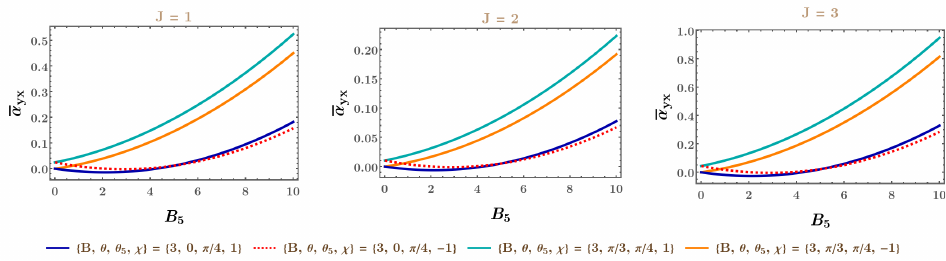}} \\
\subfigure[]{\includegraphics[width=0.95 \textwidth]{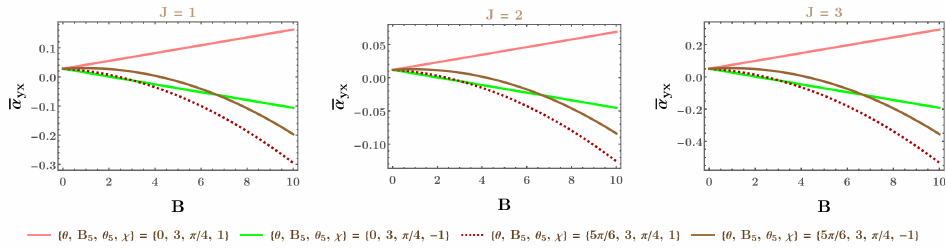}}
\caption{
	TTEC (in units of $10^{-4} $ eV) as a function of (a) $  B_5  $ (in units of eV$^2$) with $B = 3$ eV$^2$, and (b) $B$ (in units of eV$^2$) with $B_5 = 3$ eV$^2$, for various values of $\theta $ and $\theta_5 $. We have set $v_z =0.005$, $\tau = 151$ eV$^{-1}$, $\mu=0.1 $ eV, and $\beta = 1160 $ eV$^{-1}$. The curves are parabolic in each case, with the vertex of the parabola being shifted towards right(left) for $\chi = -1$($\chi =1 $), resulting in a chirality-dependent bifurcation. However, for special cases, the dependence is linear. For $ B \cos \theta = 0 $($ B \sin \theta = 0 $), the response is linear-in-$B$ as long as $ B_5 \cos \theta_5 $($ B_5 \sin \theta_5 $) is nonzero. Analogously, for $ B_5 \cos \theta_5 = 0 $($ B_5 \sin \theta_5 = 0 $), the response is linear-in-$B_5$ as long as $ B \cos \theta $($ B \sin \theta $) is nonzero, and is directly proportional to $\chi $. The pink and green lines of subfigure (b) demonstrate such a special case when $\theta =0 $.
\label{figalphaxyB}}
\end{figure}

\subsection{Transverse thermoelectric coefficient}

Since the structure of the integral for TTEC is similar to that for the PHC, we proceed in an analogous way and decompose it into four parts as follows:
\begin{align}
\bar \alpha_{y x }^{\chi} &= \alpha_{y x }^{\chi, {(1)}} + \alpha_{y x }^{\chi, {(2)}} 
+ \alpha_{y x }^{\chi, {(3)}} + \alpha_{y x }^{\chi, {(4)}}
\,, \nn
\alpha_{y x }^{\chi, {(1)}} & = 
\tau \, e  \int \frac{d^3  \mathbf k}
{( 2\, \pi)^{3}} \, D_\chi \,
v_x \, v_y \, 
\frac{  \mathcal E_\chi -\mu }  {T}\,\frac{\partial  f^{(0)}} {\partial \mathcal{E}_{\chi}} \,, \quad
\alpha_{y x }^{\chi, {(2)}} = \tau \, e^3 \,  B^{\rm tot}_{x} \, B^{\rm tot}_{y} 
\int \frac{d^3  \mathbf k}{( 2\, \pi)^{3}} \, 
D_\chi  \left  ( \boldsymbol v \cdot  \mathbf \Omega_\chi \right )^2\,
\frac{  \mathcal E_\chi -\mu }  {T}
\,\frac{\partial  f^{(0)}} {\partial \mathcal{E}_{\chi}}  \, , \nn
\alpha_{y x }^{\chi, {(3)}} & =  \tau \, e^2 \,B^{\rm tot}_{y} 
\int \frac{d^3  \mathbf k}{( 2\, \pi)^{3}} \, D_\chi \, v_x \, 
\left  ( \boldsymbol v \cdot  \mathbf \Omega_\chi \right )
\, \frac{  \mathcal E_\chi -\mu }  {T} \,
\frac{\partial  f^{(0)}} {\partial \mathcal{E}_{\chi}}  \, ,  
\quad
\alpha_{y x }^{\chi, {(4)}}
=  \tau \, e^2 \, B^{\rm tot}_x
\int \frac{d^3  \mathbf k}{( 2\, \pi)^{3}} \, D_\chi \, 
v_y  \, \left  ( \boldsymbol v \cdot  \mathbf \Omega_\chi \right ) 
\, \frac{  \mathcal E_\chi -\mu }  {T}
\, \frac{\partial  f^{(0)}} {\partial \mathcal{E}_{\chi}}  \,.
\end{align}
The integrals evaluate to
\begin{align}
\bar    \alpha_{y x }^{\chi, {(1)}} = 
\frac{ \tau \, e^3 \,J^2 \, v_z \,  \alpha_J^{\frac{2}{J}} \,  {\sqrt \pi}
\, \mu^{-1-\frac{2}{J}}\, \zeta }
{96 \,\beta } \,  
B^{\rm tot}_{x} \,  B^{\rm tot}_{y} \, 
\frac{ \Gamma (4-\frac{1}{J})}  {\Gamma (\frac{9}{2}-\frac{1}{J})}  \,, \quad   
\alpha_{y x }^{\chi, {(2)}}  =
\frac{  B^{\rm tot}_{y} }  {  B^{\rm tot}_x } \, \alpha_{xx}^{\chi, {(2)}}  \,,\quad
\alpha_{y x }^{\chi, {(3)}}  = \alpha_{y x }^{\chi, {(4)}} = 
\frac{ B^{\rm tot}_y } {2 \, B^{\rm tot}_x } \, \alpha_{xx}^{\chi, {(3)}} \,.
\end{align}

The final expression for TTEC is found to be
\begin{align}
\bar    \alpha_{y x } = 
\frac{ \tau \, e^3 \,J^2\, v_z \, \alpha_J^{\frac{2}{J}} \, 
{\sqrt \pi} \, \mu^{-1-\frac{2}{J}} \,\zeta}   
{96 \, \beta } \, 
N_{y x } \,  B^{\rm tot}_{x} \,  B^{\rm tot}_{y} \, 
\frac{ \Gamma (2 -\frac{1}{J})} {\Gamma (\frac{9}{2}-\frac{1}{J})} \,  .
\end{align}
Comparing with Eq.~\eqref{eqn:sigmaxyfinal}, we observe that $
\partial_\mu \bar \sigma_{yx}^{\chi} = - \frac  {3\, e\, \beta }  {\pi^2} \,\bar \alpha_{yx}^{\chi}
+ \mathcal{O} (\beta^{-2}) $. Therefore, once again, we find that the Mott relation $ L_{ab}^{12}  =  - \frac{\pi^2} {3\, e\, \beta } \,\partial_\mu L_{ab}^{11} $ (valid in the $\beta \rightarrow \infty $ limit) is satisfied \cite{xiao06_berry}. 
The behaviour of the TTEC for some representative parameters are shown in Figs.~\ref{figalphaxytheta} and \ref{figalphaxyB}. Fig.~\ref{figalphaxytheta} illustrates the behaviour of the TTEC as a function of $  \theta  $ for various values of $B$, $B_5$, and $\theta_5 $. Fig.~\ref{figalphaxyB} illustrates the dependence of the TTEC on $B_5$ and $B $. The overall characteristics are similar to those observed for the PHC and hence, in order to capture a slightly different parameter regime, we use a somewhat higher value of $\mu $ than that used for the PHC curves.

\section{Summary and outlook}
\label{sec_summary} 

In this paper, we have considered planar Hall (or planar thermal Hall) configurations such that a 3d Weyl or multi-Weyl semimetal is subjected to a conjunction of an electric field $\mathbf E $ (or temperature gradient $\nabla_{\mathbf r } T$) and an effective magnetic field $\mathbf B^{\rm tot}$, oriented at a generic angle with respect to each other. The $z$-axis is chosen to be along the direction along which the mWSM shows a linear-in-momentum dispersion, and is perpendicular to the plane of $\mathbf E $ (or $\nabla_{\mathbf r } T$) and $\mathbf B^{\rm tot}$. The effective magnetic field consists of two parts --- (a) an actual/physical magnetic field $\mathbf B $, and (b) an emergent magnetic field $\mathbf B_5 $ which arises if the sample is being subjected to elastic deformations (strain tensor field). Since $\mathbf B_5 $ exhibits a chiral nature, because it couples to conjugate nodal points with opposite chiralities with opposite signs, $\mathbf B^{\rm tot}$ is given by $\mathbf B + \chi \, \mathbf B_5 $. The interplay of the orientations of these two components of $\mathbf B^{\rm tot}$ with respect to the direction of the electric field (or temperature gradient) gives rise to a rich variety of possibilities in the response characteristics of the electric conductivity tensors and thermoelectric coefficients. Moreover, we have the longitudinal and transverse components of these response tensors at our disposal to obtain the signatures of the corresponding WSMs/mWSMs. We have computed the analytical expressions of these transport coefficients using a Boltzmann formalism in the limit of low magnetic fields and low temperatures, and under the relaxation time approximation for the collision integrals. Using these expressions, we have illustrated the behaviour of the response in some realistic parameter regimes. Due to the presence of the axial pseudomagnetic field, the characteristics are dependent on the chirality of the node. In addition, since the WSMs and mWSMs have different values of topological charges (quantified by $J$), the magnitude and sharpness of the conductivity tensor profiles strongly depend on the type of semimetal chosen to study. This can be understood from the explicit dependence of the analytical expressions on $J$.

Let us elaborate on the experimental evidence of the phenomena discussed in this paper. The PHE results from the nontrivial Berry phase and chiral anomaly, which is manifested by a negative magnetoresistance, a quadratic-in-magnetic-field dependence of the magnetoconductance, and an oscillatory behaviour with the angle between the electric and the (actual) magnetic fields. These have been measured in numerous experiments --- a few examples involve materials like ZrTe$_{5}$ \cite{li_2016}, TaAs \cite{checng-long}, NbP and NbAs \cite{li_nmr17}, and Co$_3$Sn$_2$S$_2$ \cite{shama}, hosting Weyl semimetallic bandstructures. On the other hand, experimental set-ups for the controlled application of strain gradients, leading to the realization of artificial gauge fields, have been demonstrated in Ref.~\cite{diaz_2022}. Such fabrications are still in the nascent stage, and more needs to be done in order to achieve controlled set-ups that can realize $ \mathbf B_5 $ in the presence of a nonzero $\mathbf B$. Nonetheless, our results provide a concrete prediction of what to expect in such experimental situations.

In the future, it will be worthwhile to improve the characterization of the response tensors by including a momentum/energy-dependent relaxation time $\tau$ and also by incorporating internode scatterings \cite{girish2023}. Furthermore, for connecting with realistic scenarios, one would be interested to study similar transport properties in the presence of disorder and/or many-body interactions \cite{ips-seb, ips_cpge, ips-biref, ips-klaus, rahul-sid, ipsita-rahul-qbt, ips-qbt-sc}. Yet another possible direction is to apply an additional time-periodic drive \cite{ips-sandip, ips-sandip-sajid, ips-serena, *ips-mwsm-floquet} on the system, for example, by shining circularly polarized light. Last, but not the least, under the influence of a strong quantizing magnetic field, when Landau level formation cannot be ignored, the fingerprints of the thermoelectric coefficients are extremely relevant \cite{ips-kush,fu22_thermoelectric,staalhammar20_magneto,munoz_torsion-ll,yadav23_magneto}.

\section*{Acknowledgments}

RG is grateful to Ram Ramaswamy for providing the funding to complete this paper.


\appendix

\section{Obtaining conductivity tensors using the semiclassical Boltzmann formalism} 
\label{secboltz}


In this appendix, we review the semiclassical Boltzmann formalism \cite{mermin,ips-kush-review}, which we have used to determine the transport coefficients. In the presence of a magnetic field, the semiclassical Boltzmann transport approach works well for small magnetic fields and small cyclotron frequency, i.e., in the regime where the Landau level quantization can be ignored.

For a system in three spatial dimensions, we define the distribution function (alternatively, the probability density function) $ f_n( \mathbf r , \mathbf k, t) $ for the Bloch band (labelled by the index $n$) with the crystal momentum $\mathbf k$ and dispersion $\epsilon_n(\mathbf k)$, such that
\begin{align}
dN_n = g_n \,f_n( \mathbf r , \mathbf k, t) \,
\frac{ d^3 \mathbf k}{(2\, \pi)^3 } \,d^3 \mathbf r
\end{align}
is the number of particles in an infinitesimal phase space volume $
dV_p = \frac{ d^3 \mathbf k}{(2\, \pi)^3 } 
\,d^3 \mathbf r $, centered at $\left \lbrace \mathbf r , \mathbf k \right \rbrace $ at time $t$. Here $g_n$ is the degeneracy of the band. If we neglect the orbital magnetization of the Bloch wavepackets as well as the contributions from the spin-orbit interactions, and assume that the Bloch bands are topologically trivial, the Hamilton's equations of motion for Bloch electrons in electromagnetic fields are given by (cf. Chapter--12 of Ref.~\cite{mermin}):
\begin{align}
\label{eqkin2}
{\hbar} \, \dot{\mathbf r} = \partial_{\mathbf k} \epsilon_n ({\mathbf k})\,,
\quad 
\hbar\,\dot{\mathbf k}
= {\mathcal Q} \left( {\mathbf E} +  
\frac{ \dot{\mathbf r} \cross {\mathbf B} } {c}
\right ) ,
\end{align}
where $\mathcal Q $ is the electric charge of a single quasiparticle, and $\mathbf E $ and $\mathbf B $ are the externally applied electric applied electric and magnetic fields. We have denoted the total time derivatives by the widely used convention of overhead dots. This leads to the kinetic equation
\begin{align}
\label{eqkin3}
\left [ \partial_t  
+ {\boldsymbol v}_n  \cdot \nabla_{\mathbf r} 
+ \frac{\mathcal Q} {\hbar} \left(
\mathbf E + \frac{ {\boldsymbol v}_n   \cross {\mathbf B} } {c} \right) 
\cdot \nabla_{\mathbf k} \right ] f_n ( \mathbf r , \mathbf k, t)
= \left[ \frac{\partial f_n ( \mathbf r , \mathbf k, t)}
{\partial t} \right]_{\text{coll}} \,,
\end{align}
where 
\begin{align}
\boldsymbol{v}_n({\mathbf{k}})
= \frac{1}{\hbar} \, \nabla_{\mathbf k} \epsilon_n ({\mathbf k }) 
\end{align}
is the Bloch velocity (or group velocity). The the right-hand-side contains the correction term $ I_{\rm{coll}} \equiv
\left[ \frac{\partial f_n( \mathbf r , \mathbf k, t) }
{\partial t} \right]_{\text{coll}}$, also known as the collision integral. $ I_{\rm{coll}}$ arises due to collisions of the quasiparticles, as the name suggests, and has to be added as a perturbation to correct the Liouville's equation in the presence of scattering events.

Here we employ a simple model for the collision integral which is widely known as the
relaxation time approximation. Let us denote the static distribution
function of the quasiparticles as
\begin{align}
\label{eqdist}
f^{(0)}_n (\mathbf r,\mathbf k) 
\equiv 
f^{(0)}_n \big (\epsilon_n(\mathbf k) , \mu (\mathbf r), T (\mathbf r) \big )
= \frac{1}
{e^{ \frac{\epsilon_n  (\mathbf k )-\mu  (\mathbf r )} 
		{k_B \, T  (\mathbf r )}} + 1}\,,
\end{align} 
which describes a local equilibrium situation at
the subsystem centred at position $\mathbf r$, at the local temperature $T(\mathbf r )$, and with the local chemical potential $\mu (\mathbf r )$. Now we make the ansatz
\begin{align}
I_{\text{coll}}= -
\frac{ f_n(\mathbf r,\mathbf k, t)
	-f^{(0)}_n (\mathbf r,\mathbf k)
}
{\tau (\mathbf k)} \,,
\end{align}
where $\tau (\mathbf k) $ is called the relaxation time and is, in general, a function of the momentum. Physically, $\tau (\mathbf k) $ represents the characteristic time scale within which the system relaxes to equilibrium, after the occurrence of a scattering process relevant for the problem under consideration.

In any system, the quasiparticles transport thermal energy (i.e., heat) simultaneously with electric charge. Hence, in a generic situation, we need to consider a sample subjected to weakly-spatially-varying temperature $T (\mathbf r)$ and chemical potential $\mu(\mathbf r) $. It is convenient to introduce a combined electrochemical potential and the generalized (external) force field defined by
\begin{align}
\eta(\mathbf r) = \Phi(\mathbf r) 
- \frac{\mu (\mathbf r) } {\mathcal Q}
\text{ and }
\boldsymbol{\mathcal E} (\mathbf r) = 
-\nabla_{\mathbf r} \eta (\mathbf r) \,,
\end{align}
respectively, where $\Phi (\mathbf r) $ is the electrostatic potential such that $\mathbf E = -\nabla_{\mathbf r } \Phi $. Hence, Eqs.~\eqref{eqkin2} and \eqref{eqkin3} must be generalized to
\begin{align}
\label{eqkin21}
{\hbar} \, \dot{\mathbf r} = \partial_{\mathbf k} \epsilon_n ({\mathbf k})\,,
\quad 
\hbar\,\dot{\mathbf k}
= {\mathcal Q} \left( \boldsymbol{\mathcal E} +  
\frac{ \dot{\mathbf r} \cross {\mathbf B} } {c}
\right )
\end{align}
and
\begin{align}
\label{eqkin32}
& \left [
\partial_t  
+ {\boldsymbol v}_n  
\cdot \nabla_{\mathbf r} 
+ \frac{\mathcal Q} {\hbar} \left(  \boldsymbol{\mathcal E}
+ \frac{ {\boldsymbol v}_n   \cross {\mathbf B} } {c}
\right) 
\cdot \nabla_{\mathbf k} \right ] f_n
=
\frac{ f^{(0)}_n (\mathbf r,\mathbf k)
	- f_n(\mathbf r,\mathbf k, t) }
{\tau (\mathbf k)} \,,
\end{align}
respectively.

In the presence of a nontrivial topological charge in the bandstructure, the Boltzmann equation of Eq.~\eqref{eqkin3} will get modified. We specifically focus on 3d nodal-point semimetals with nonzero Chern numbers. Considering transport for a single node of chirality $\chi$, Eq.~\eqref{eqkin21} will get modified to \cite{Sundurum:1999,son13_chiral}
\begin{align}
\label{eqrkberry}
{\hbar}   \, \dot{\mathbf r} = 
\partial_{\mathbf k} \epsilon_n  (\mathbf k)
- 
\hbar\, \dot{\mathbf k} \cross \mathbf{\Omega }_{\chi ,n}\,,\quad 
\hbar\,\dot{\mathbf k}
= {\mathcal Q} \left(\boldsymbol{\mathcal E} +  
\frac{ \dot{\mathbf r} \cross {\mathbf B} } {c}
\right ) ,
\end{align}
where $\mathbf{\Omega }_{\chi ,n} (\mathbf k)$ is the Berry curvature of the node, which is a pseudovector
expressed by
\begin{align}
\mathbf{\Omega }_{\chi ,n} (\mathbf k) = 
i \left \langle  \nabla_{\mathbf k}
u^{\chi}_n  (\mathbf k) \right | \cross \left | \nabla_{\mathbf k}  u_n^{\chi}  (\mathbf k) \right \rangle .
\end{align} 
The Berry  curvature arises from the Berry phases generated by $|u_n^{\chi} (\mathbf k)\rangle$, where $\lbrace |u_n^{\chi} (\mathbf k)\rangle \rbrace $ denotes the set of orthonormal Bloch cell eigenstates for the one-particle Hamiltonian $H_\chi (\mathbf k)$ representing the low-energy effective description of the node with band energies $\lbrace \epsilon_n \rbrace $.
It can be checked that $\mathbf{\Omega }_{\chi ,n} $  is proportional to $\chi$ and, hence, it has opposite signs for an energy band with index $n$ at nodes of opposite chiralities.
Here we will neglect the orbital magnetization of the Bloch wavepackets.

The two coupled equations in Eq.~\eqref{eqrkberry} can be solved to obtain
\begin{align}
\label{eqrkberry1}
\dot{\mathbf{r}} &=  D_{\chi, n} 
\left[ 
{ \boldsymbol{v}}_n
- \frac{ \mathcal Q } {\hbar}
\, \boldsymbol{\mathcal E} \cross \mathbf{\Omega}_{\chi, n}
-
\frac{\mathcal Q} { \hbar \, c} 
\left(   \mathbf{\Omega }_{\chi ,n} \cdot {\boldsymbol{v}}_n  \right)
\mathbf{B}\right], \quad
\hbar \, \mathbf{\dot{k}} =  
D_{\chi, n}   \,
\mathcal Q 
\left[  \boldsymbol{\mathcal E}
+\frac{ {\boldsymbol{v}}_n
	\cross \mathbf{B}}  {c} 
- \frac{\mathcal Q  } { \hbar \, c} \,
\left ( \boldsymbol{\mathcal E}  \cdot {\mathbf B} \right )
\mathbf{\Omega }_{\chi ,n} \right],
\end{align}
where
\begin{align}
D_{\chi,n}^{-1} = 1- \frac {\mathcal Q} {\hbar\, c }
\, {\mathbf B} \cdot \mathbf{\Omega }_{\chi ,n}  \,.
\end{align} 
The physical significance of the function $D_{\chi,n} $ can be understood as follows.
While studying the semiclassical dynamics of Bloch electrons, Xiao \textit{et al.}~\cite{prl_niu} observed that the Liouville's theorem on the conservation of phase space volume element $ dV_p $ is violated by the Berry phase. This breakdown of the Liouville’s theorem is remedied by introducing a modified density of states in the phase space such that the number of states in the volume element $D_{\chi,n}^{-1} \, dV_p$ remains conserved. Based on this modification, the classical phase-space probability density is now given by \cite{son13_chiral,prl_niu,duval06_Berry,Son:2012}
\begin{align}
F_n (\mathbf r, \mathbf k, t) = D_{\chi,n}^{-1} (\mathbf r, \mathbf k) \, 
f_n( \mathbf r , \mathbf k, t) \,.
\end{align}
Probability conservation implies that, in the absence of collisions, $F_n$ satisfies the continuity equation in the phase space, viz., $\frac{ d F_n} {dt} = 0 $.
Incorporating all these ingredients, Eq.~\eqref{eqkin32} should be modified to \cite{lundgren14_thermoelectric,amit_magneto}
\begin{align}
\label{eqkin33}
& D_{\chi, n}  \,
\left [ \partial_t  
+ \left \lbrace 
{ \boldsymbol{v}}_n
-\frac{ \mathcal Q } {\hbar} 
\,\boldsymbol{\mathcal E} \cross \mathbf{\Omega}_{\chi,n}
-
\frac{\mathcal Q} {\hbar \, c} \left(   \mathbf{\Omega }_{\chi ,n} \cdot 
{\boldsymbol{v}}_n  \right)
\mathbf{B} \right \rbrace 
\cdot \nabla_{\mathbf r} 
+ \frac{\mathcal Q} {\hbar}
\left(  \boldsymbol{\mathcal E}
+ \frac{ {\boldsymbol v}_n   \cross {\mathbf B} } {c}
\right) 
\cdot \nabla_{\mathbf k} 
- \frac{\mathcal Q^2 } {\hbar^2 \,c} \,
\left ( \boldsymbol{\mathcal E}  \cdot {\mathbf B} \right )
\mathbf{\Omega }_{\chi ,n}   \cdot \nabla_{\mathbf k} 
\right ] f_n
\nn & \, =   
\frac{ f^{(0)}_n (\mathbf r,\mathbf k)
	- f_n(\mathbf r,\mathbf k, t) }
{\tau (\mathbf k)} \,.
\end{align}
For the sake of simplicity, here we have assumed that only intranode scatterings are relevant in contributing to $\tau$, thus ignoring the internode scattering processes.

In order to obtain a solution to the full Boltzmann equation for small time-independent values of $\mathbf E$, $\nabla_{\mathbf r} \mu$, and $\nabla_{\mathbf r} T$, we assume a slight deviation $\delta  f_n(\mathbf r,\mathbf k)$ from the equilibrium distribution of the quasiparticles, which does not have any explicit time-dependence. Hence, the non-equilibrium time-independent distribution function is given by
\begin{align}
f_n(\mathbf r,\mathbf k, t)\equiv  f_n(\mathbf r,\mathbf k)
=  f^{(0)}_n(\mathbf r,\mathbf k) +  \delta  f_n(\mathbf r,\mathbf k)\,.
\end{align} 
The magnetic field, however, is not assumed to be small.
It is reasonable to have assumed the solution $ \delta  f_n$ not to have any explicit time-dependence since the applied fields and gradients are time-independent.
The gradients of the equilibrium distribution function $ f^{(0)}_n$ evaluate to
\begin{align}
\nabla_{\mathbf r}  f^{(0)}_n (\mathbf r,\mathbf k) 
& = 
\left( \nabla_{\mathbf r} \mu
+ \frac{ \epsilon_n - \mu} {T} 
\, \nabla_{\mathbf r} T \right )
\left( - \frac{\partial  f^{(0)}_n } {\partial  \epsilon_n } \right ) \text{ and }
\nabla_{\mathbf k}  f^{(0)}_n (\mathbf r,\mathbf k) 
= \hbar \,   {\boldsymbol v}_n 
\, \frac{\partial  f^{(0)}_n (\mathbf r,\mathbf k)} {\partial  \epsilon_n } \,.
\end{align}
Let us consider a uniform chemical potential such that $ \nabla_{\mathbf r} \mu =0 $.
We assume that all of the quantities, viz., $\mathbf E $, $\nabla_{\mathbf r} T$, and the resulting $\delta f_n$, are of the same order of smallness. The spatial gradient $\nabla_{\mathbf r}  f^{(0)}_n$ is parallel to the thermal gradient $\nabla_{\mathbf r} T$, and we consider situations where $\mathbf E $ and  $\nabla_{\mathbf r} T$ are applied along the same direction. Hence, the term $\mathcal Q \left (  {\mathbf E} \cross \mathbf{\Omega}^{\chi }_n \right) \cdot
\nabla_{\mathbf r} f^{(0)}_n $ in Eq.~\eqref{eqkin33} gives zero.

To the leading order in the ``smallness parameter'', the so-called \textit{linearized Boltzmann equation} is given by
\begin{align}
\label{eqkin5}
& \left [
\left \lbrace {\boldsymbol{v}}_n 
-  \frac{\mathcal{Q}} {\hbar \, c} \left(
{\mathbf \Omega}_{\chi, n} \cdot
{\boldsymbol{v}}_n   \right)  \mathbf B \right \rbrace
\cdot \left( -
\frac{  \epsilon_n - \mu } {T}  \, \nabla_{\mathbf r} T 
+ {\mathcal Q} \, \mathbf E
\right ) \right] 
\frac{\partial  f^{(0)}_n (\epsilon_n , \mu, T) } {\partial \epsilon_n }
- \frac{   
	Q \, {\mathbf B} \cdot
	\left( {\boldsymbol v}_n  \cross \nabla_{\mathbf k}
	\right) } 
{ \hbar \, c} \, \delta f_n (\mathbf r,\mathbf k)
\nonumber \\
& \quad  = -\frac{\delta f_n (\mathbf r,\mathbf k)} 
{D_{\chi, n}  \,\tau(\mathbf k) } \,.
\end{align}
In our linearized approximation, the term $\mathcal Q \left (  {\mathbf E} \cross \mathbf{\Omega}_{\chi, n} \right) \cdot
\nabla_{\mathbf r} \delta f_n $ from Eq.~\eqref{eqkin33} does not contribute, as it is of second order in smallness. To solve the above equation, we need to make an appropriate ansatz, as outlined in (a) Refs.~\cite{nandy_2017_chiral,amit_magneto} for planar Hall effect; and (b) Ref.~\cite{nandy_thermal_hall} for planar thermal Hall effect.

Let the contributions to the average electric and thermal currents from the quasiparticles, associated with the node being considered, be ${\mathbf J}^\chi$ and ${\mathbf J}^{Q,\chi}$. The response matrix, which relates the resulting generalized currents to the driving electric potential gradient or temperature gradient, can be expressed as
\begin{align}
\label{eqcur1}
\begin{pmatrix}
	J_a^\chi \vspace{0.2 cm} \\
	{J}_a^{Q,\chi} 
\end{pmatrix} & = 
\sum \limits_b
\begin{pmatrix}
	L_{a b }^{11} & L_{a b }^{12} 
	\vspace{0.2 cm}  \\
	L_{a b }^{21} & L_{a b }^{22}
\end{pmatrix}
\begin{pmatrix}
	\mathcal{E}_b
	\vspace{0.2 cm}  \\
	-{ \partial_{r^a} T } 
\end{pmatrix} ,
\end{align}
where $ \lbrace a, b \rbrace  \in \lbrace x,\, y, \, z \rbrace $ indicates the Cartesian components of the current vectors and the response tensors in 3d. The set $ [L_{\chi}] \equiv \lbrace L^{11}_{ab}, \,  L^{12}_{ab}, \, L^{21}_{ab} , \, L^{22}_{ab} \rbrace $ represents the transport coefficients.
Using the solutions for $ f_n = f_n^{(0)} + \delta f_n $, we get their explicit expressions~\cite{nandy_2017_chiral, nandy_thermal_hall,amit_magneto}. The components of the magnetoelectric conductivity tensor $\sigma$, thermopower tensor $ S $ (also known as the Seebeck coefficient), Peltier coefficient $\Pi$, and magnetothermal conductivity tensor $\lambda $ can be extracted from $[L_\chi]$ as follows \cite{mermin,ips-kush-review}:
\begin{align}
\label{eq:kappa}
& \sigma_{ a b}^\chi  =  L_{ a b}^{11} \,  ,
\quad
S_{ a b}^\chi = \sum \limits_{ a^\prime}
\left(L^{11}\right)^{-1}_{ a   a^\prime }
L_{ a^\prime b}^{12} \, , \quad
\Pi_{ a b} ^\chi= \sum \limits_{ a^\prime}
L_{ a   a^\prime}^{21}   
\, \left(L^{11}\right)^{-1}_{ a^\prime b} \,,\quad 
\lambda_{a b}^\chi =
L_{ a b }^{22}
- \sum \limits_{ a^\prime, \, b^\prime }
L_{ a  a^\prime }^{21}
\left(L^{11}\right)^{-1}_{  a^\prime  b^\prime }
L_{ b^\prime b }^{12}  \,.
\end{align}

The dc charge current density~\cite{nandy_2017_chiral,amit_magneto} and the thermal current density~\cite{nandy_thermal_hall,lundgren14_thermoelectric} take the forms
\begin{align}
\label{eqcur}
{\mathbf J}^\chi
& =   {\mathcal Q} \, \sum_n g_n  \int
\frac{ d^3 \mathbf k}{(2\, \pi)^3 } \,
D_{\chi, n}^{-1}   \, \dot{\mathbf r}
\,  f_n( \mathbf r , \mathbf k) \text{ and} \nn
\mathbf{J}^{Q,\chi}
& = \sum_n g_n  \int
\frac{ d^3 \mathbf k}{(2\, \pi)^3 } 
\,D_{\chi, n}^{-1}   
\, \dot{\mathbf r} 
\left( \epsilon_n - \mu \right)  f_n( \mathbf r , \mathbf k)\,,
\end{align}
respectively.

Let us consider the magnetic field to be applied in the $xy$-plane, such that $\mathbf B = B \left( \cos \theta \,{\mathbf{\hat x}} + \sin \theta \, {\mathbf{\hat y}}\right)$. An electric field $\mathbf E = E \,  {\mathbf{\hat x}} $ is applied in a coplanar set-up, with $\nabla_{\mathbf r} T = 0 $. In this paper, we have considered only the case of a momentum-independent $\tau$. From the solutions obtained in Refs.~\cite{nandy_2017_chiral,amit_magneto}, and setting ${\mathcal Q} = - e $ (where $e$ is the magnitude of the charge of an electron) and $g_n= 1$ (ignoring the degeneracy due to electron's spin), we arrive at the following expressions for a single band of chirality $\chi$ and band index $n = s$:
\begin{align}
\label{eqsigmatot}
& \sigma_{ a b }^\chi 
= \sigma_{ a b }^{\chi,\, \rm AHE}
+ \sigma_{ a b }^{\chi,\, \mathbf \Gamma}
+ \bar \sigma_{ a b }^\chi \,, \quad
\sigma_{ a b }^{\chi,\, \rm AHE}
= -\frac{e^2} {\hbar } \,\epsilon_{a b c} 
\int \frac{ d^3 \mathbf k}{(2\, \pi)^3 } \, \Omega_{\chi, s}^c \,  f^{(0)}_s \,,\nn
&
\bar \sigma_{ a b }^\chi 
=- e^2 \, \tau
\int \frac{ d^3 \mathbf k}{(2\, \pi)^3 } \, D_{\chi, s} 
\left [ {v_s}_a  +\frac{e\, B_a} {\hbar \, c} \left( 
{\boldsymbol{v}}_s \cdot \mathbf \Omega_{\chi, s} \right)
\right ]
\left [ { v_s}_b  +\frac{e\,  B_b  } {\hbar \, c} \left( 
{\boldsymbol{v}}_s \cdot \mathbf \Omega_{\chi, s} \right)
\right ]
\, \frac{\partial  f^{(0)}_s } {\partial  \epsilon_s } \,.
\end{align}
Here $\sigma_{ a b }^{\chi,\, \rm AHE}$ represents the ``intrinsic anomalous'' Hall effect \cite{haldane,pallab_axionic,burkov_intrinsic_hall} (which is, evidently, completely independent of the scattering rate), $ \sigma_{ a b }^{\chi,\, \mathbf \Gamma}$ is the  Lorentz-force contribution to the conductivity, and $ \bar \sigma_{ a b }^\chi $ is the Berry-curvature-related conductivity coefficient.
For a momentum-independent $\tau$, $ \sigma_{ a b }^{\chi,\, \mathbf \Gamma}$ is much smaller than the other terms~\cite{nandy_2017_chiral} and it has only a nonzero transverse component involving $zx$ --- hence, we neglect it. Furthermore, we are not interested in $\sigma_{ a b }^{\chi,\, \rm AHE}$, as it turns out to be zero for the continuum model we have taken here.

Next we consider a magnetic field $\mathbf B = B \left( \cos \theta \,{\mathbf{\hat x}} + \sin \theta \, {\mathbf{\hat y}}\right)$ and a temperature gradient $\nabla_{\mathbf r}  T = \partial_x T\, {\mathbf{\hat x}} $, with $\mathbf E$ is set to zero. We are interested in finding the form of the thermopower tensor, for the same semimetallic node described above, which is given by $S^\chi_{ab}$. For this, we need to evaluate the thermoelectric coefficient $L_{ab}^{12}$, which we denote by $\alpha_{ a b }^\chi $.
Using the solutions described in Refs.~\cite{pal22b_berry, nandy_thermal_hall}, we get the expressions
\begin{align}
\label{eqalphatot}
& \alpha_{ a b }^\chi 
=  \alpha_{ a b }^{\chi,\, \rm AHE}
+  \alpha_{ a b }^{\chi,\, \mathbf \Gamma}
+ \bar \alpha_{ a b }^\chi \,, \quad
\alpha_{ a b }^{\chi,\, \rm AHE}
= 
\frac{e} {\hbar } \,\epsilon_{a b c} 
\int \frac{ d^3 \mathbf k}{(2\, \pi)^3 } \,
 \Omega_{\chi, s}^c \, 
\, \frac{\left( \epsilon_s - \mu \right)
} {T} \, f^{(0)}_s \,,\nn
& \bar \alpha_{ a b }^\chi 
= e \, \tau
\int \frac{ d^3 \mathbf k}{(2\, \pi)^3 } \, D_{\chi, s} 
\left [ {v_s}_a  +\frac{e\, B_a} {\hbar \, c} \left( 
{\boldsymbol v}_s \cdot \mathbf \Omega_{\chi, s} \right)
\right ]
\left [ {v_s}_b  +\frac{e\,  B_b } {\hbar \, c} \left( 
{\boldsymbol{v}}_s \cdot \mathbf \Omega_{\chi, s} \right)
\right ]
\, \frac{\left( \epsilon_s - \mu \right) } {T}
\, \frac{\partial  f^{(0)}_s } {\partial  \epsilon_s } \,.
\end{align}
Analogous to the earlier case, $ \alpha _{ a b }^{\chi,\, \rm AHE}$ arises independent of an external magnetic field, $ \alpha_{ a b }^{\chi,\, \mathbf \Gamma}$ results from the Lorentz-force-like contributions, and $ \bar \alpha_{ a b }^\chi $ is the Berry-curvature-related part. We ignore the first two contributions in this paper, as $\alpha _{ a b }^{\chi,\, \rm AHE}$ vanishes, while $\alpha_{ a b }^{\chi,\, \mathbf \Gamma}$ has a subleading contribution for a momentum-independent $\tau$ (with the only nonzero component being $zx$).

To summarize, in this paper, we will consider the behaviour of the parts $\bar \sigma_{ a b }^\chi$ and $\bar \alpha_{ a b }^\chi$.

\section{Strain-induced pseudomagnetic field} 
\label{secstrain}

In this section, we  review how mechanical strain, which induces elastic deformations in a material, can be modelled as pseudogauge fields. We take Weyl semimetal as an example, and focus on the effects of a torsion, which gives rise to an axial pseudomagnetic field.

Following the treatment in Ref.~\cite{pikulin_gauge}, we consider a specific model describing the low-energy degrees of freedom in 3d Dirac semimetals (e.g., Cd$_3$As$_2$ and Na$_3$Bi).  Near the $ \Gamma $-point of the Brillouin zone, the dispersion can be described by the four-band effective continuum Hamiltonian 
\begin{align}
\label{h2}
& H_D({\mathbf k})=\epsilon_0({\mathbf k})+
\begin{pmatrix}
	M_{\mathbf k} & A \, k_- & 0 & 0\\
	A \, k_+& -M_{\mathbf k} & 0 & 0 \\
	0 & 0 & -M_{\mathbf k} &-A \, k_- \\
	0 & 0 & -A \, k_+ & M_{\mathbf k}
\end{pmatrix},  \nonumber \\
& \epsilon_0(\mathbf k)=C_0 + C_1\, k_z^2 + 
C_2 \left (k_x^2+k_y^2 \right ), \quad 
M_{\mathbf k}= M_0 + M_1 \, k_z^2 + M_2 
\left (k_x^2+k_y^2 \right ) ,\quad
k_\pm=k_x \pm i \,k_y \,.
\end{align}
The constants $ \lbrace C_0, \, C_1, \, C_2\rbrace $, $A$, and $  \lbrace M_0, \, M_1, \, M_2\rbrace$ are material-dependent parameters which are obtained from the $\mathbf k \cdot \mathbf p$ expansion of the first-principles-calculations \cite{wang_dirac,wang_dirac2}. The spectrum of the model harbours a pair of Dirac points located at 
\begin{align} 
\label{eqnQ}
{\mathbf K}_\eta = \lbrace 0,0,\eta \,Q \rbrace \,,
\quad  Q= \sqrt{-M_0/M_1}\,, \nonumber
\end{align}
where $\eta=\pm$ is the valley index.

The above Hamiltonian, on being regularized on a lattice, takes the form:
\begin{align}
H^{\text{latt}}_D 
= \epsilon_0^{\rm{latt}} (\mathbf k)
+ \begin{pmatrix}
	h^{\rm{latt}} & 0 \\
	0 & - h^{\rm{latt}}
\end{pmatrix}, 
\quad \epsilon_0^{\rm{latt}}
({\mathbf k}) =c_0 + c_1 \cos (a\,k_z) +
c_2 \left[ \cos (a\,k_x) + \cos (a\,k_y) \right ], 
\end{align}
\begin{align}
& h^{\rm latt}( {\mathbf k} )
= m_ {\mathbf k} \, \sigma_z +
\Lambda \left[ \,
\sigma_x \sin ( a \,k_x) 
+ \sigma_y \sin (a \, k_y ) \,\right] ,
\quad
m_ {\mathbf k} =t_0 +
t_1\cos (a \, k_z )  + 
t_2 \left[ \, \cos (a \, k_x ) 
+ \cos (a \, k_y ) \, \right ], 
\nonumber \\ &
t_0=M_0+2\,(M_1+2 \,M_2)/a^2\,, 
\quad t_{1} = -2 \,M_{1}/a^2\,, \quad 
t_{2} = -2 \,M_{2}/a^2\, , \quad 
\Lambda=A/a \,,
\end{align}
where $a$ is the lattice constant.
We note that the $2\times 2$ Hamiltonian $h^{\rm latt}
( {\mathbf k} )$ describes a single pair of Weyl nodes at the points $ {\mathbf K}_\eta =
\left \lbrace 0, \, 0,\, \eta \, Q  \right \rbrace  $,
where $ \cos(a \,Q)= - \frac{t_0+2 \,t_2 } {t_1} $.
In the vicinity of each node, we can expand  $h^{\rm latt}( {\mathbf K}_\pm + {\mathbf q})$ in $ {\mathbf q}$ to obtain the Weyl Hamiltonians
\begin{align} 
\label{h3c}
{\tilde h}_{\eta}( {\mathbf q})
= \hbar \sum\limits_{a=x, y, z} 
v^{W,\eta}_a \,
\sigma_a \,  q_a \,,
\quad
{\boldsymbol{v}}^{W,\eta} =
\hbar^{-1} \, a \, \left \lbrace \Lambda, \Lambda,
-\eta \, t_1 \sin(a\,Q) \right \rbrace .
\end{align}
The chirality of the corresponding Weyl node is given by
\begin{align} \label{h3d}
\chi_{\eta}=
\text{sgn}( v_x^{W,\eta} \,v_y^{W,\eta} \, 
v_z^{W,\eta} ) = -\eta.
\end{align}

In order to incorporate the effects of elastic strain, we modify the Hamiltonian $ h^{\rm latt}( {\mathbf k} )$ by replacing the hopping amplitude along the $z$-direction as
\begin{align} 
\label{h4}   
t_1\, \sigma_z \rightarrow t_1 \left (1-u_{zz} \right ) \sigma_z + i \, \Lambda \left(  u_{zx}\, \sigma_x  + u_{zy}
\, \sigma_y \right ), \quad
\end{align}
where $ u_{ab} =
\frac{\partial_{r^a} \, u^d_b
\, +  \, \partial_{r^b} \, u^d_a} {2}
$ is the symmetrized strain tensor, and $\boldsymbol{u}^d $ represents the displacement vector. The elastic distortion, expressed through Eq.~\eqref{h4}, generates additional terms of the form  
\begin{equation}
\label{h5}
\delta h^{\rm latt}( {\mathbf k} )
= -t_1\, u_{zz} \, \sigma_z \cos(a\,k_z)
+\Lambda \left (u_{xz} \, \sigma_x -u_{yz} \, 
\sigma_y \right ) \sin  (a \, k_z) \,.
\end{equation}
With this perturbation added, the position of a node shifts to
$ {\mathbf K}^{\rm str}_\eta = \mathbf{K}_\eta - \frac{\eta\, q } {c} {\boldsymbol{\mathcal A}}$. The shift vector ${\boldsymbol{\mathcal A}}$,  given by
\begin{align} 
{\boldsymbol{\mathcal A} } =
-{\hbar \,  c\over e \, a}
\left \lbrace u_{xz} 
\sin (a\,Q), \,  u_{yz} \sin (a\,Q),
\, u_{zz} \cot  (a \, Q)
\right \rbrace ,
\end{align}
can be thought of an emergent effective vector gauge potential.

Expanding in the vicinity of $ {\mathbf K}^{\rm str}_\eta $, the linearized Hamiltonian of the strained crystal is captured by
\begin{align} \label{h6}
{\tilde h}_{\eta} ^{\rm str} ( {\mathbf q}) =
\hbar \sum\limits_{a=x, y, z} 
v^{W,\eta}_a \,
\sigma_a \left(q_a
- { \eta \,e\over c} \,{\mathcal A}_a \right),
\end{align}
For $a \, Q\ll 1$ (continuum limit), we may approximate $ \sin (a\,Q)\simeq a\,Q\simeq a\sqrt{-M_0/M_1}$ and $\cot(a\,Q) \simeq 1/(a\,Q)$, leading to
\begin{align} \label{h7}
{\boldsymbol{\mathcal A} } =
-{\hbar \,  c\over e }
\left \lbrace u_{xz} \, Q,
\, u_{yz} \, Q, \, \frac{u_{zz}} {a^2 \, Q}
\right \rbrace .
\end{align}
The above form shows that in a Weyl semimetal, with nodes located along the $k_z$-component of the momentum, the $u_{a z}$ component of the strain field tensor acts on the low-energy fermionic excitations as a gauge potential ${\boldsymbol{\mathcal A} }$. It behaves as a chiral gauge field because it couples to the quasiparticles around the two nodes with opposite signs. The resulting pseudomagnetic field is given by
\begin{align}
\mathbf B_5  = \nabla_{\mathbf r} \cross {\boldsymbol{\mathcal A} }\,.
\end{align}

\bibliography{ref_strain}

\end{document}